\documentclass[%
 reprint,
superscriptaddress,
 amsmath,amssymb,
 aps,
 prx
floatfix,
]{revtex4-2}

\usepackage[displaymath]{lineno}  
\usepackage{graphicx}
\usepackage{amssymb}
\usepackage{amsmath}
\usepackage[utf8]{inputenc}
\usepackage{cancel}
\usepackage[usenames,dvipsnames]{color}
\usepackage{psfrag}
\usepackage{epsfig}
\usepackage{bbm}
\usepackage{bm}
\usepackage{dsfont}
\usepackage{soul}
\usepackage{colonequals}
\usepackage{float}
\usepackage{enumitem}
\usepackage{appendix}
\usepackage{lmodern}
\usepackage[T1]{fontenc}
\usepackage{physics}                                                
\usepackage{hyperref}
\usepackage[capitalize]{cleveref}
\usepackage{xcolor}			
\usepackage{siunitx} 
\usepackage{flafter}

\DeclareSIUnit{\molar}{M}

\newcommand{\ccite}[1]{\IfSubStr{#1}{,}{Refs.~}{Ref.~}\cite{#1}} 

\newcommand{\tf}{t_\mathrm{f}}

\newcommand{\HM}{\textbf{H}}

\definecolor{DarkRed}{RGB}{153,0,0}
\definecolor{DarkGreen}{RGB}{0,153,0}
\definecolor{DarkBlue}{RGB}{0,0,153}

\newcommand{\out}[1]{ }

\newcommand{\dt}{\mathrm{d}t}

\newcommand{\la}{\left\langle}
\newcommand{\ra}{\right\rangle}

\newcommand{\rmd}{\mathrm{d}}

\newcommand{\f}{\bm}
\newcommand{\fx}{{\f R}}
\newcommand{\fr}{{\f r}}
\newcommand{\fl}{{\f \lambda}}
\newcommand{\He}{{H}}
\newcommand{\Dx}{{\Delta x}}

\graphicspath{{Figures/}}
\begin{document}

\nolinenumbers

\preprint{APS/123-QED}

\title{Energy-Efficient Control of Interacting Microscopic Systems:\\
When Longer Paths Save Energy}

\author{Samuel Monter}\email{samuel.monter@uni-konstanz.de}
\affiliation{Faculty of Physics, University of Konstanz, Konstanz, Germany}
\author{Lars T. Stutzer}\email{lars.stutzer@ds.mpg.de}
\affiliation{Max Planck Institute for Dynamics and Self-Organization, Goettingen, Germany}
\author{Sarah A. M. Loos}\email{sarah.loos@ds.mpg.de}
\affiliation{Max Planck Institute for Dynamics and Self-Organization, Goettingen, Germany}
\affiliation{Department of Applied Mathematics and Theoretical Physics, University of Cambridge, UK}
\author{Clemens Bechinger}\email{clemens.bechinger@uni-konstanz.de}
\affiliation{Faculty of Physics, University of Konstanz, Konstanz, Germany}
\affiliation{Centre for the Advanced Study of Collective Behaviour, University of Konstanz, Germany
}

\date{\today}

\begin{abstract}
We experimentally and theoretically study the thermodynamically optimal control of interacting multiple-particle systems, focusing on collections of colloidal particles individually confined in optical traps. We investigate protocols that transport the system between prescribed trap configurations within a fixed time in the most energy efficient way. For Markovian systems with conservative pairwise interactions, we establish a general result in the low-noise limit: optimal particle trajectories are linear in space and time, corresponding to steady straight-line motion, irrespective of the specific interaction potential, even for nonlinear forces. Thus,
conservative interactions do not modify the geometry of the optimal paths. This property breaks down in the presence of strong noise or nonconservative interactions. For the paradigmatic case of hydrodynamic coupling, we demonstrate experimentally that optimal control can involve curved trajectories that significantly reduce the energetic cost by exploiting collectively generated fluid flows. The emergence of curved paths as optimal solutions highlights a fundamental distinction between non-interacting and interacting systems and reveals a cooperative mechanism for energy-efficient control.
\end{abstract}

\maketitle


The controlled manipulation of microscopic systems far from equilibrium is a central theoretical and experimental challenge, with direct relevance to  microrobotics~\cite{soto2020medical,chi2024perspective,ju2025technology} and biological molecular motors~\cite{toyabe2011thermodynamic,mugnai2020theoretical,gupta2022optimal}. At these length scales, thermal fluctuations and viscous dissipation dominate the dynamics, motivating a strong interest in control strategies that steer system behavior while minimizing energetic cost. This objective is naturally formulated within stochastic thermodynamics~\cite{sekimoto1998langevin,seifert2012stochastic}.\\
\indent Previous studies have largely focused on single-particle systems~\cite{optimal2007schmiedl,alvarado2025optimal}, including extensions to active particles~\cite{gupta2023efficient,davis2024active,garcia2025optimal,schuttler2025active}, environments with memory~\cite{loos2024universal}, nonlinear forces~\cite{zulkowski2015optimal,sivak2016thermodynamic,zhong2022limited,gupta2022optimal,engel2023optimal,oikawa2025experimentally}, and nonequilibrium state transitions~\cite{zulkowski2013optimal,monter2025optimal,olsen2025harnessing}.
Remarkably, even though only a single degree of freedom is controlled in these settings, the corresponding optimal protocols already exhibit rich and nontrivial features when minimizing work input or heat dissipation.\\ 
\indent By contrast, much less is known about the optimal control of interacting many-particle systems. Here, high dimensionality of the state space and the emergence of interaction-induced collective effects pose new fundamental challenges, and severely complicate the extension of single-particle control strategies to genuine many-body dynamics. Recent approaches have begun to address this problem using coarse-grained field-theoretic descriptions~\cite{davis2024active,soriani2025control}. However, a systematic understanding of optimal control strategies at the level of interacting particles remains largely unexplored.\\
\indent Using a combination of analytical, numerical, and experimental approaches, we study the optimal control of multiple interacting colloidal particles driven by movable optical traps. The control parameters are the trap positions, which can be adjusted individually and are translated between prescribed initial and final configurations within a fixed time. The particle positions constitute the system’s physical degrees of freedom and evolve dynamically in response to trap motion, thermal fluctuations, and particle interactions. The objective is to minimize the total work performed during this transport, thereby extending earlier single-particle studies \cite{optimal2007schmiedl} to multi-particle systems. Trapping each particle individually, allows us to resolve the structure of optimal collective dynamics arising from both conservative and nonconservative interactions.\\
\indent For Markovian systems with conservative pairwise interactions and harmonic driving potentials, we establish analytically in the low noise limit that optimal particle trajectories are universally linear in space and time, minimizing dissipation irrespective of the specific interaction potential, even in the presence of nonlinear interaction forces. Deviations from this behavior arise only when thermal noise becomes sufficiently strong. Using a perturbative approach, we show that interaction-induced nonlinearities generate corrections at second order in the noise amplitude. In contrast, nonconservative interactions alter the geometry of optimal trajectories already in the zero-noise limit. For the case of hydrodynamic coupling, which is always present in viscous environments, we demonstrate experimentally that optimal control involves curved particle trajectories that reduce the total work through momentum transfer via the surrounding fluid. As a result, groups of particles can benefit energetically from moving cooperatively, reminiscent of collective motion in schools of fish.

\begin{figure}
    \centering
    \includegraphics[width=1\linewidth]{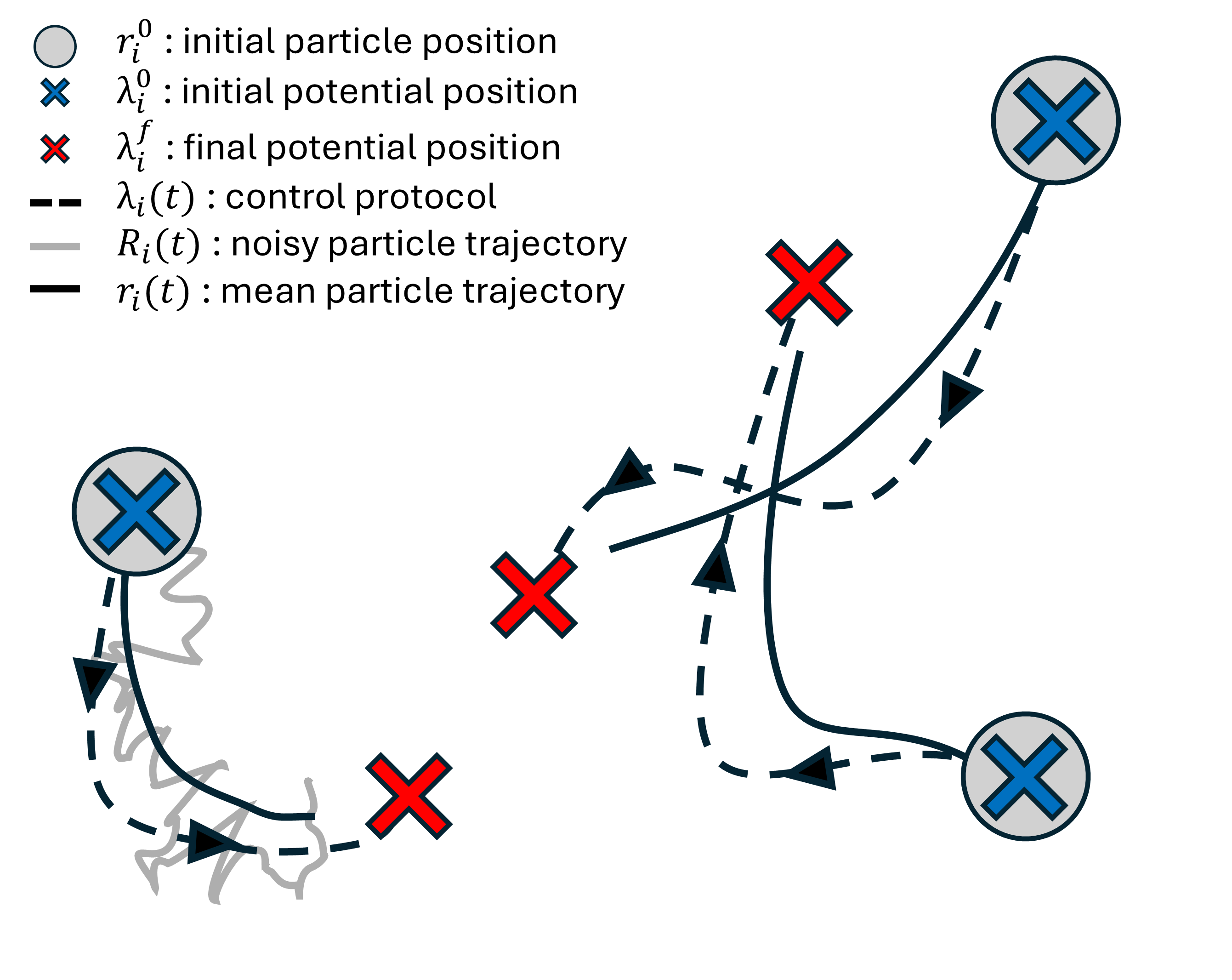}
    \caption{Sketch of the optimal control problem. Starting from an initial configuration with trap centers (blue~$\times$) and particle positions (gray~$\bigcirc$) at $\lambda_i^0$ and $r_i^0$, respectively, the control protocol $\lambda_i(t)$ (dashed black lines) translates the traps to their final positions $\lambda_i^\mathrm{f}$ (red~$\times$) within a finite time $\tf$. The particles position evolves according to the traps driving and thermal fluctuations (gray line); averaging over many realizations yields the mean particle trajectories (black lines). At $t=\tf$, the particle positions generally do not coincide with $\lambda_i^f$ but subsequently relax to the new equilibrium positions without additional work.}
    \label{fig:sketch}
\end{figure}


\section{Control problem and notation}
We consider $N$ particles with positions $\fx_i\in\mathds{R}^d$, $i=1, 2, \dots, N$ in spatial dimension $d\geq 1$, which interact with each other via pairwise interactions (specified below).
Each particle is confined in a harmonic potential $V_\mathrm{trap}= \frac{\kappa}{2}\sum_{i=1}^N \norm{\fx_i-\fl_i}^2$ with stiffness $\kappa$ and center $\fl_i\in\mathds{R}^d$, where $\norm{\cdot}$ denotes the Euclidean norm measuring distances.\\
\indent The $N$ trap centers constitute our $N\,d$ control parameters. The control objective is to move the traps from prescribed initial positions [$\fl_i(t=0)=\fl_i^0$] to final positions [$\fl_i(t=t_\mathrm{f})=\fl_i^\mathrm{f}$], see Fig.~\ref{fig:sketch}, within a prescribed amount of time $t_\mathrm{f}$ while minimizing the total mean work \cite{sekimoto1998langevin} input
\begin{align}
    W = 
    \left\langle \int_0^{t_\mathrm{f}}  \rmd s\, \sum_{i=1}^N\frac{\partial V_\mathrm{trap}}{\partial {\fl}_i}\dot{\fl}_i(s)\right\rangle
  \, .\label{Work}
\end{align}

\section{Conservative interactions}

\begin{figure*}
    \centering
    \includegraphics[width=.9\linewidth]{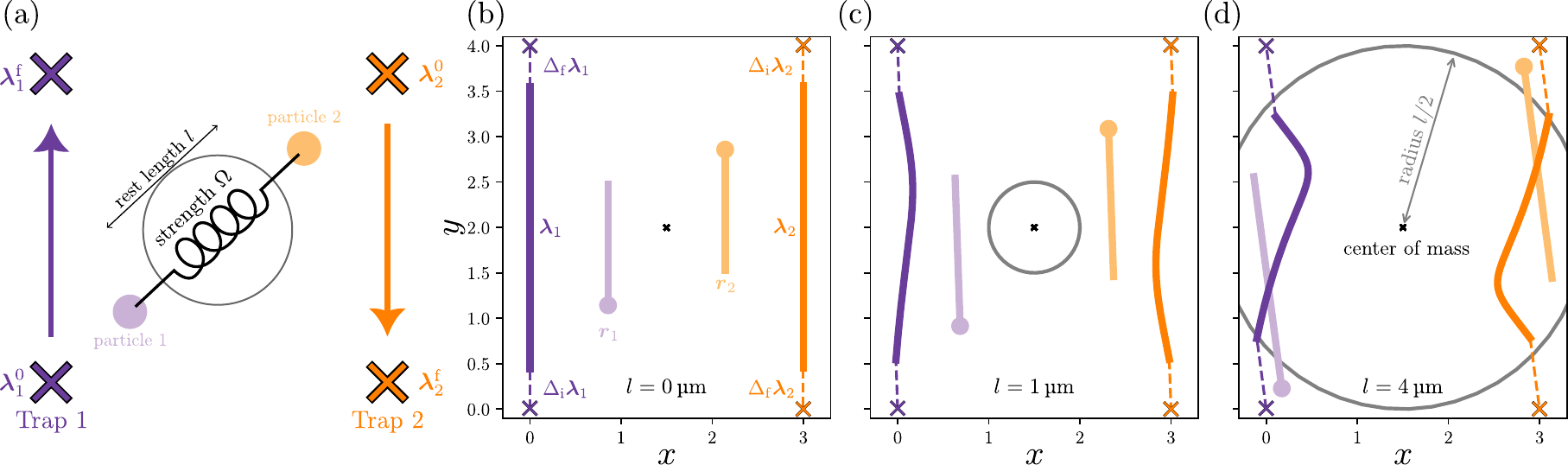}
    \caption{Example control problem with conservative particle interactions.
    (a) Sketch of the setup: two optical traps ($\{\f \lambda_i\}$), each containing a particle ($\{\f r_i\}$, circles), are driven anti-parallel from prescribed initial positions 
    to final positions (crosses). 
    The particles are coupled by a harmonic spring with rest length $l$ and stiffness $\Omega$. The gray circle of diameter $l$ marks the cross-over at which the spring force changes sign. 
    (b-d) Optimal solutions for different rest lengths $l$. 
    In all cases, the optimal protocols exhibit jumps (dashed lines) at the beginning ($\Delta_\mathrm{i}\f \lambda_i$) and end ($\Delta_\mathrm{f}\f \lambda_i$). The initial mean particle positions are obtained from the equilibrium distribution corresponding to the initial trap configuration. The black dot marks the symmetry center and center of mass of the system. 
    As predicted by the general theoretical argument, the mean particle trajectories are straight lines in all cases, while the optimal protocols clearly deviate from straight-line motion for $l>0~\unit{\micro\meter}$. These curved protocols balance the nonlinear elastic coupling between the colloids.
    A finite noise strength (here zero) would not affect the solutions for $l=0~\unit{\micro\meter}$ (panel b), but for $l>0~\unit{\micro\meter}$ (panel c and d) it may induce nonlinear particle trajectories. Parameters: final time ${t_\mathrm{f}}=1~\unit{\second}$, trap coupling $\kappa=3~\unit{\micro\newton\per\meter}$, spring coupling $\Omega=2~\unit{\micro\newton\per\meter}$, vertical displacement of traps $H=4~\unit{\micro\meter}$, horizontal distance traps $\Dx=3~\unit{\micro\meter}$, and friction constant $\gamma=1~\unit{\milli\pascal\second}$.
    }
    \label{fig:SpringLengthComparison}
\end{figure*}

We first address the case of pairwise conservative interactions, $V_\mathrm{int} = \sum_{i>j}V_\mathrm{int}^{ij}(\fx_i-\fx_j)$, ubiquitous in physical systems from electrostatic interactions to elastic and gravitational forces. 
For the sake of generality, here we account for inertial effects as well as additional external fields $V_\mathrm{ext}= \sum_{i=1}^N V_\mathrm{ext}^i(\fx_i)$, and allow for arbitrary dimensions and particle numbers.
The dynamics are assumed to follow a Langevin equation of the general structure, $m_i\ddot{\fx}_i = -\f\gamma_i\dot{\fx}_i - \nabla_i V[\lambda] + \sqrt{2k_\mathrm{B}T\f\gamma_i}\f\zeta_i$,
where the total potential decomposes as $V=V_\mathrm{trap}+V_\mathrm{int}+V_\mathrm{ext}$, $m_i$ is the mass of the colloids, $\f\gamma_i$ denotes 
a constant and symmetric, positive-definite friction matrix, and $\f\zeta$ is an additive Gaussian white noise term with $\langle \f\zeta_i\rangle=\f0$, $\langle [\f\zeta_i(t)]_n[\f\zeta_j(s)]_m\rangle = \delta_{ij}\delta_{mn}\delta(t-s)$, and temperature $T$.\\
\indent General insights into the solutions of the control problem can be deduced by rewriting the work in Eq.~\eqref{Work} as a functional of the mean particle positions $\fr_i=\langle \fx_i\rangle$. To this end, we insert the Langevin equation into Eq.~\eqref{Work}, and identify boundary contributions from the energy difference between the initial and final states of the protocol (see Appendix~\ref{AppendixConservative}). In general, this leads to expressions of the work functional and associated Lagrangian of the optimization that are not analytically tractable due to nonlinearities. The notable exceptions are cases where (i) all potentials are harmonic 
or (ii) the system operates in the zero-noise limit $T\to 0$.
If (i) or (ii) are satisfied, we obtain the following form 
\begin{align}
    W &= \Delta V + \Delta E_\mathrm{kin} + \int_0^{t_\mathrm{f}}\rmd s \sum_{i=1}^N\dot{\fr}_i^T\f\gamma_i\dot{\fr}_i\label{conservativeW}\,,
\end{align}
with
\begin{align}
    \Delta V &= V_\mathrm{trap}+V_\mathrm{int}+V_\mathrm{ext}\Big|_0^{t_\mathrm{f}}\,,\\
    \Delta E_\mathrm{kin} &= \frac{1}{2}\sum_{i=1}^Nm_i\norm{\dot{\fr}_i(s)}^2\Big|_0^{t_\mathrm{f}}.
\end{align}

\noindent Equation~\eqref{conservativeW} shows that conservative forces contribute exclusively as boundary terms. The only path-dependent (bulk) contribution arises from frictional dissipation. Since the latter is minimized by solutions with $\{\ddot{\fr_i}=0\}$, we conclude that, for linear or zero-noise nonlinear systems, optimal mean particle trajectories exhibit a straight-line motion with constant velocity (see Appendix \ref{AppendixConservative} for details).\\
\indent While optimal particle trajectories are linear in space and time, the same does not generally hold for the control protocols. In fact, the very need to maintain straight-line motion of the particle can, in general, force the control protocol itself to become curved and discontinuous. Mathematically, this is seen by the fact that $\f \lambda^*_i = \f g_i(\{\f r^*_i\})$ is generally given by a nonlinear functional $\f g_i$ due to the gradient forces $\nabla_i (V_\mathrm{int} + V_\mathrm{ext})$.\\
\indent To illustrate this finding, we consider a system of two particles coupled elastically via a harmonic spring with potential $V_\mathrm{int} = \Omega/2 \left(\norm{\fr_1-\fr_2}-l\right)^2$ and stiffness $\Omega$. If the spring has zero rest length $l=0$, the associated control problem is fully linear. Consistent with our general argument  [scenario (i)], optimal mean particle trajectories generally follow straight-lines for any noise level (see App.~\ref{AppendixExampleSpring}). We find that the corresponding control protocols are linear in the bulk and exhibit jumps at the beginning and end, analogous to the corresponding single-particle problem~\cite{optimal2007schmiedl}.
By contrast, a finite spring rest length $l>0$ is sufficient to render the problem nonlinear, whenever the elastic forces and driving directions are not parallel. We calculate in Appendix~\ref{AppendixExampleSpring} the optimal solutions at $T=0$ [scenario (ii)] for dragging two particles in opposite directions , as sketched in  Fig.~\ref{fig:SpringLengthComparison}(a). The nonlinearities arising from finite spring rest length are sufficient to generate curved optimal protocols, see Fig.~\ref{fig:SpringLengthComparison}, while mean particle trajectories remain straight-lines, confirming our general argument for scenario (ii).\\
\indent Beyond the scenarios (i) and (ii), the combination of nonlinear interaction forces and finite noise amplitude generally renders the problem of finding optimal protocols analytically intractable. 
Remarkably, however, the linearity property of optimal trajectories extends beyond the deterministic limit to the regime of weak noise.
To show this, we expand in Eq.~\eqref{Work} the stochastic positions  $\fx_i = \fx_{i,0} + \sigma\fx_{i,1} + \sigma^2\fx_{i,2}+\dots$ in noise amplitude $\sigma\propto\sqrt{T}$.
Restricting to optimal solutions, the optimal protocol can be expanded analogously, $\f\lambda_i^* = \fl^*_{i,0} + \sigma\fl^*_{i,1} + \sigma^2\fl^*_{i,2}+\dots$. Up to first order in $\sigma$, the work retains the same structure as in Eq.~\eqref{conservativeW}, 
\begin{align}\label{eq:W-lownoise}
    W = \Delta V + \Delta E_\mathrm{kin} + \int_0^{t_\mathrm{f}}\rmd s \sum_{i=1}^N(\dot{\fr}_{i,0}^*)^T\f\gamma_i\dot{\fr}_{i,0}^*+\mathcal{O}(\sigma^2)\,.
\end{align}
Here, ${\fr}_{i,k}^*$ is the optimal average trajectory of the $i$th particle at $k$th order in $\sigma$. Equation~\eqref{eq:W-lownoise} shows that corrections arising from the interplay between nonlinear interaction forces and thermal fluctuations enter only at second order, $\sigma^2\propto T$. Consequentially, in the low-noise limit, even nonlinear systems are expected to maintain the straight-line optimal solutions.
See App.~\ref{noiseCorrection} for a more thorough discussion.\\ 
\indent 
We conclude that, in conservative athermal systems, optimal mean particle trajectories follow straight lines, whereas nonlinear systems are expected to exhibit deviations once noise exceeds the small-noise regime. Since a sharp theoretical criterion for the validity of this approximation is difficult to obtain, controlled experiments are needed to determine whether the small-noise limit applies in a given system.\\ 
\indent While the theoretical argument aligns with general intuition for conservative systems, we emphasize that this result is highly nontrivial. For instance, in more constrained settings with fewer control parameters than physical degrees of freedom, optimal solutions become nonlinear—even in the presence of purely conservative and linear interactions (see Ref.~\cite{loos2024universal} for an example).

\section{Experimental Results for Non-conservative Interactions}

\begin{figure*}
	\centering
	\includegraphics[width=0.9\textwidth]{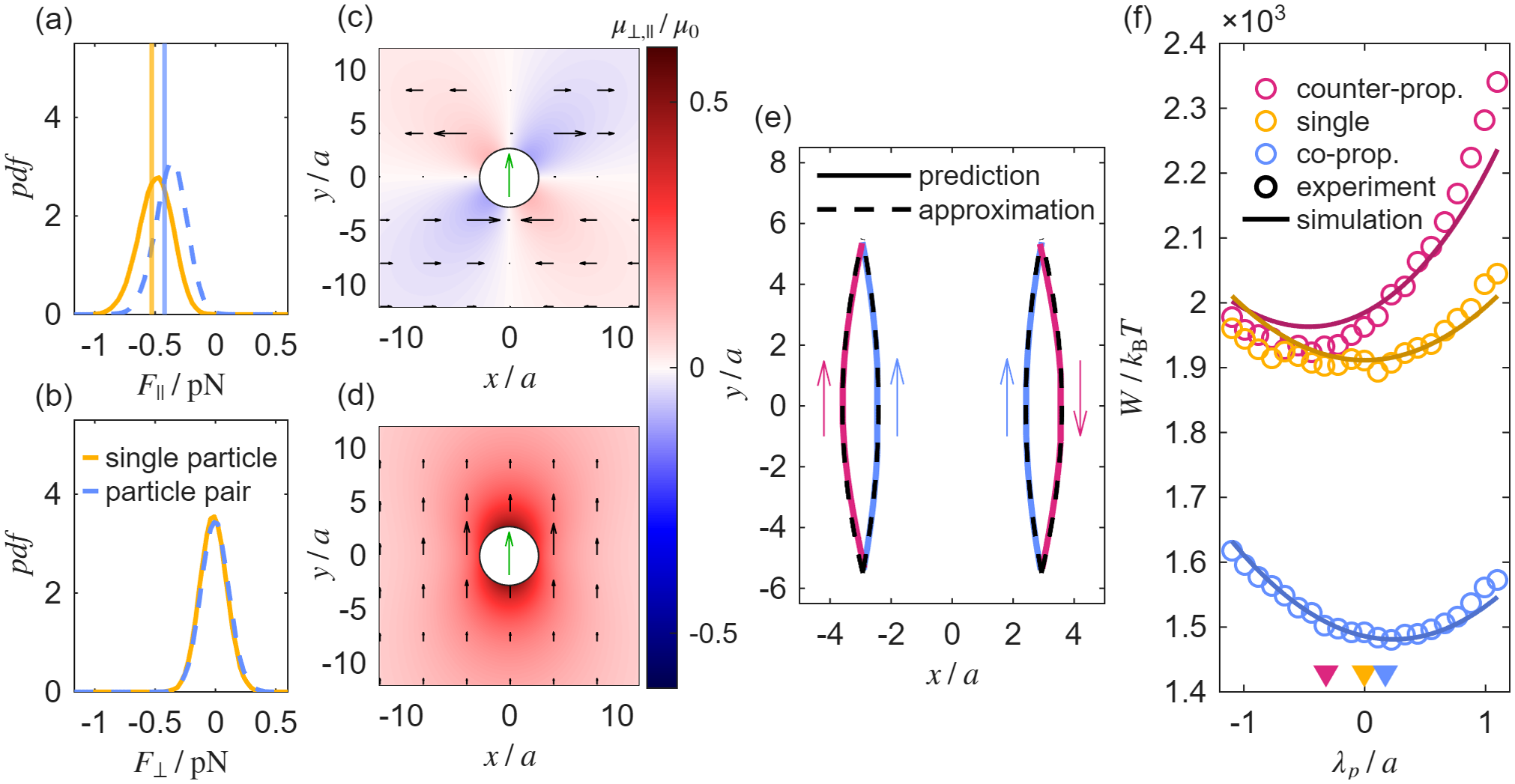} 
	\caption{
    (a,b) Mean force required to drag colloidal particles at constant velocity
    $v=3~\unit{\micro\meter\per\second}$ ($a=1.37~\unit{\micro\meter}$).
    The longitudinal force $F_\parallel$ is reduced when a second particle is dragged
    alongside at a separation of $8~\unit{\micro\meter}$, while the transverse force
    $F_\perp$ has zero mean. Vertical lines indicate theoretical predictions based on a
    fluid viscosity of $6.8~\unit{\milli\pascal\second}$
    \cite{cheng2008formula,volk2018density}. Systematic deviations are attributed to
    particle-size variations and residual spatial inhomogeneities of the trap stiffness.   
    (c,d) Color coded spatial dependence of the longitudinal and transverse components
    $\mu_{\parallel,\perp}$ of the simplified mobility tensor $\mathbf H$
    [see Eq.~\eqref{eq:Hsimple}]. A force applied to the central particle (green arrow) induces a velocity of the second particle (black arrows). The reduction of $F_\parallel$ in (a) originates from longitudinal hydrodynamic coupling $\mu_\parallel$, whereas
    $\mu_\perp$ does not affect head-on motion, explaining the vanishing $F_\perp$.
    (e) Predicted optimal control protocols for two traps
    ($\kappa=3~\unit{\micro\newton\per\meter}$, $a=1.37~\unit{\micro\meter}$,
    $t_\mathrm{f}=5~\unit{\second}$). Blue: co-propagating configuration; red:
    counter-propagating configuration. Dashed lines show parabolic approximations
    parameterized by $\lambda_p$ [see Eq.~\eqref{eq:lambda_parab}].
    (f) Measured average work $ W$ as a function of $\lambda_p$
    (open symbols), compared to simulations (solid lines). For a single particle, the
    straight path ($\lambda_p=0$) minimizes work. For co-propagating particles, the
    minimum shifts to $\lambda_p\simeq0.23\,a$ due to hydrodynamic force transfer,
    whereas for antiparallel driving $\lambda_p\simeq$ 
    $-0.44~a$ (the measured minima are marked as symbols at the bottom). Quantitative agreement is obtained by fitting the viscosity
    ($\eta_\rightrightarrows=6.0$, $\eta_{\leftrightarrows}=6.2$, $\eta_{\rightarrow}=6.9~
    \unit{\milli\pascal\second}$), consistent with literature values. Residual
    discrepancies are attributed to particle polydispersity ($\sim4\%$),
    trap-stiffness variations ($\sim5\%$), and weak laser-induced heating.
     }
	\label{fig:expres2d}
\end{figure*}

The general result established in the preceding section relies on the fact that all interparticle forces are conservative, so that their contribution to the work reduces to boundary terms. This simplification no longer applies in the presence of non-conservative forces which cannot be derived from a scalar potential and which arise in many physical contexts, including velocity-dependent Lorentz or Coriolis forces, and dissipative or active forces that break action–reaction symmetry. Non-conservative interactions give rise to additional path-dependent contributions in the work functional, Eq.~\eqref{Work}. As a consequence, neither temporal nor geometric linearity of the optimal motion can be generally expected, and the structure of the optimal trajectories will depend sensitively on the specific form of the non-conservative interactions.\\
\indent As a particular relevant example of non-conservative forces, we now experimentally investigate the influence of hydrodynamic interactions on the optimal control problem discussed above. Hydrodynamic interactions are ubiquitous for particles moving in viscous liquids and originate from the velocity-dependent flow fields generated when externally driven particles move through the solvent. The imposed particle motion continuously dissipates energy into the fluid while simultaneously inducing flow fields that transport momentum and influence the motion of neighboring particles. As a result, the coupling mediated by a dissipative medium renders hydrodynamic forces intrinsically non-conservative.\\
\indent In our experiments, we realize a minimal system consisting of two colloidal particles suspended in a viscous fluid and confined by optical tweezers. The particles are silica spheres with a radius of $a = 1.37 \pm 0.06~\unit{\micro\meter}$. Externally prescribed time-dependent protocols are applied to the optical trap centers in the sample plane, while the particle positions evolve dynamically in response to the moving traps and the resulting hydrodynamic interactions. The optical traps are generated via time sharing using a dual-axis acousto-optical deflector operated at an update rate of $10~\unit{\kilo\hertz}$~\cite{jones2015optical,gieseler2021optical}, providing harmonic confinement and precise, independent control of the trap motion in the focal plane of the objective. Particle positions are recorded with a spatial resolution of $6~\unit{\nano\meter}$ at a frame rate of $100~\unit{\hertz}$. To enable a quantitative comparison with theory, we restrict the particle interaction to hydrodynamic coupling and suppress additional interactions with the confining walls. Consequently, the particles are at least $\geq30~\unit{\micro\meter}$ away from all sample boundaries, minimizing both direct particle–wall interactions and wall-induced hydrodynamic corrections~\cite{homqvist2007colloidal}. The hydrodynamic coupling between the particles is further enhanced by employing a 1:1 (volume ratio) water–glycerol mixture with a viscosity of $6.9~\unit{\milli\pascal\second}$ at the experimental temperature $T=25~\unit{\celsius}$, as described in Refs.~\cite{cheng2008formula,volk2018density}. Further details on the experimental setup and calibration measurements are provided in App.~\ref{sec:expmeth}.\\
\indent To quantify the strength of hydrodynamic coupling in our system, we first perform a reference experiment in which two particles are translated through the fluid while monitoring the forces required to maintain a prescribed velocity. The particles are separated by $8~\unit{\micro\meter} = 5.9~a$ orthogonal to the driving direction and driven in parallel at a constant velocity of $3~\unit{\micro\meter\per\second}$. The force in the driving direction $F_\parallel$ is reduced compared to that required to drag a single particle at the same velocity, as shown in Fig.~\ref{fig:expres2d}(a). By contrast, the force component orthogonal to the driving direction $F_\perp$ is indistinguishable from the single-particle reference [see Fig.~\ref{fig:expres2d}(b)], indicating that no additional interactions between particles contribute appreciably to the dynamics.\\
\indent The observed force reduction is well described by hydrodynamic interactions captured by the Rotne–Prager–Yamakawa (RPY) approximation~\cite{rotne1969variational,yamakawa1970transport,dhont1996introduction}. 
Within this framework, the constant mobility $\f\mu = \f\gamma^{-1}$ appearing in the single-particle Langevin equation is replaced by a position-dependent mobility tensor $\HM$ for interacting particles. From a fluid-mechanical perspective, this tensor provides a compact description of how forces applied to one particle generate flow fields that entrain other particles. It thereby quantifies how forces are redistributed between particles through the surrounding fluid. For two particles in two spatial dimensions, a simplified form of the RPY mobility tensor reads
\begin{equation}
	\HM = 
	\begin{pmatrix} 
		\mu_0             &0                   &\mu^{\parallel}   &\mu^{\perp}\\
		0               &\mu_0                 &\mu^{\perp}   &\mu^{\parallel}\\ 
		\mu^{\parallel}  &\mu^{\perp}       &\mu_0             &0\\ 
		\mu^{\perp}   &\mu^{\parallel}       &0               &\mu_0\\\end{pmatrix}
     \label{eq:Hsimple}
\end{equation}
where $\mu_0$ denotes the bulk mobility, while $\mu^{\parallel}$ and $\mu^{\perp}$ describe longitudinal and transverse hydrodynamic coupling between the particles (for derivation see App.~\ref{sec:HI}). The distinction between longitudinal and transverse hydrodynamic coupling directly reflects the geometry of the flow field generated by a moving particle. Flow components along the direction of motion tend to entrain neighboring particles and assist or hinder their translation depending on whether the particle co- or counter-propagate. The transverse flow components primarily induce lateral deflections without contributing to motion along the driving direction. Importantly, the coupling coefficients $\mu^{\parallel}$ and $\mu^{\perp}$ depend on the relative particle configuration [see Fig.~\ref{fig:expres2d}(c,d)], while $\mu_0$ is constant.
We calculate the mean force required to drag a single particle or a pair of particles using the RPY model with zero noise and compare it to the data measured in the reference experiment.
The only free parameter we need for the calculation is the fluid viscosity, which we take from literature~\cite{cheng2008formula,volk2018density}. The predicted forces are shown as vertical lines in Fig.~\ref{fig:expres2d}(a). For both single particles or pairs, the model slightly overestimates the force. This discrepancy can be attributed to small deviations of the tabulated viscosity and to the residual spatial variations in the optical trap stiffness (see App.~\ref{sec:setupNcali}). In addition, the model neglects thermal noise, which may contribute weak corrections.
This reference experiment provides a quantitative characterization of hydrodynamic coupling in our setup and forms the basis for the optimal-control experiments discussed below.\\
\indent To further validate our ansatz, we performed additional numerical test simulations based on the RPY-model with and without noise, in the experimentally accessible parameter regimes, dragging particle pairs at realistic separations and along different relative directions. The averaged stochastic trajectories agreed well with the corresponding deterministic trajectories, indicating that noise-induced deviations are within the error bars. We therefore employ the deterministic PRY-model throughout the remainder of the study (see App.~\ref{sec:HInoise}).\\
\indent To realize optimal control in experiments, we first require a theoretical prediction of the optimal protocol, which serves as a starting point around which the experimental parameters can be systematically varied.
Due to the nonlinear nature of the hydrodynamic interactions, we resort to numerical tools based on the optimal control Hamiltonian formalism to solve the optimization problem.
To validate our optimization approach, we compute optimal control protocols for widely separated particles (distance $>100 a$) and compare them with the known single-particle solution~\cite{optimal2007schmiedl} (for details see App.~\ref{sec:numerics}).\\
\indent Figure~\ref{fig:expres2d}(e) shows theoretically predicted optimal control protocols for two specific geometries, both involving two optical traps. 
In both configurations, the start and end positions of the traps lie on two parallel lines separated by $8\,\unit{\micro\meter}=5.9\,a$ along the $x$-axis. The traps are displaced by $15\,\unit{\micro\meter}=11.0\,a$ along the $y$-direction. In the co-propagating configuration (blue), both traps move in the same direction, i.e., start and end at identical $y$-positions. In the counter-propagating configuration (red), the start and end positions of one trap are exchanged, resulting in opposite motion along $y$. In both cases, the protocol duration, $\tf = 5~\unit{\second}$, is much longer than the particle relaxation within the trap, $\tau \approx 0.04~\unit{\second}$.
The numerically obtained optimal protocols have a curved shape which is well approximated by parabolas with a linear time-dependence of the trap velocity in $x$- and a constant trap velocity in $y$-direction 
\begin{equation}
	\f\lambda(t) = \begin{pmatrix}
		-\frac{\lambda_p}{\tf^2} (2t-t_\mathrm{f})^2 + \lambda_p\\ 
		v_y t\\
		\frac{\lambda_p}{\tf^2} (2t-t_\mathrm{f})^2 + \lambda_p\\
		\pm v_y t\\
	\end{pmatrix} +\f\lambda^0.
    \label{eq:lambda_parab}
\end{equation}
Note that $\lambda_p>0$ corresponds to protocols bending inwards. The velocity $v_y$ is set in accordance with the boundary conditions. This expression allows us to experimentally test theoretical predictions by varying a single parameter, namely the parabola depth $\lambda_p$.\\
\indent We implement these protocols experimentally for both relative driving directions as well as for a single trapped particle as a reference. The resulting average work $W$ per particle is shown in Fig.~\ref{fig:expres2d}(f). As expected for a single particle (yellow), the straight-line protocol ($\lambda_p=0$) minimizes the work. By contrast, in both two-particle configurations, the minimum of $W$ is offset from $\lambda_p=0$. In agreement with our predictions, the optimal curvature satisfies $\lambda_p^* > 0$ in the co-propagating configuration and $\lambda_p^* < 0$ in the counter-propagating configuration.
Most importantly and opposed to the conservative interacting particles (see Fig.~\ref{fig:SpringLengthComparison}) the corresponding mean particles trajectories are then no longer straight lines, but also follow a curved path, thereby increasing the traveled distance towards the target (see Fig.~\ref{fig:meantraj}). Noteworthy, the curved protocols observed here do not arise from direct attractive or repulsive forces between particles.\\
\indent To understand this surprising result, it is important to recall the hydrodynamic coupling as illustrated in Fig.~\ref{fig:expres2d}. A particle moving through a fluid creates a surrounding flow field. In the case of a single particle, the energy injected into the fluid is entirely lost due to viscous friction. For multiple particles with decreasing particle distances, the surrounding fluid transmits an increasing fraction of the force exerted through the optical traps between the particles. 
As a direct consequence, counter-propagating configurations generally require more work than co-propagating or single particle configurations.
Curved protocols allow the particles to transiently access configurations in which fluid-mediated forces reduce particle dissipation, even at the expense of longer geometric paths.
The co- and counter-propagating configurations represent two complementary manifestations of this mechanism. For parallel driving, hydrodynamically induced flows assist collective motion, favoring inward-bending trajectories that temporarily reduce the distance between particles. In contrast, for antiparallel driving, the induced flows oppose motion, increasing resistance and leading to outward-bending optimal trajectories. Such detours, however, come at a cost: deviations from  a straight line increase the total path length and, therefore, viscous dissipation. Curved trajectories arise from a competition between the energetic gain from hydrodynamic momentum transfer and the additional dissipation caused by the detour. \\
\begin{figure}
    \centering
    \includegraphics[width=0.9\linewidth]{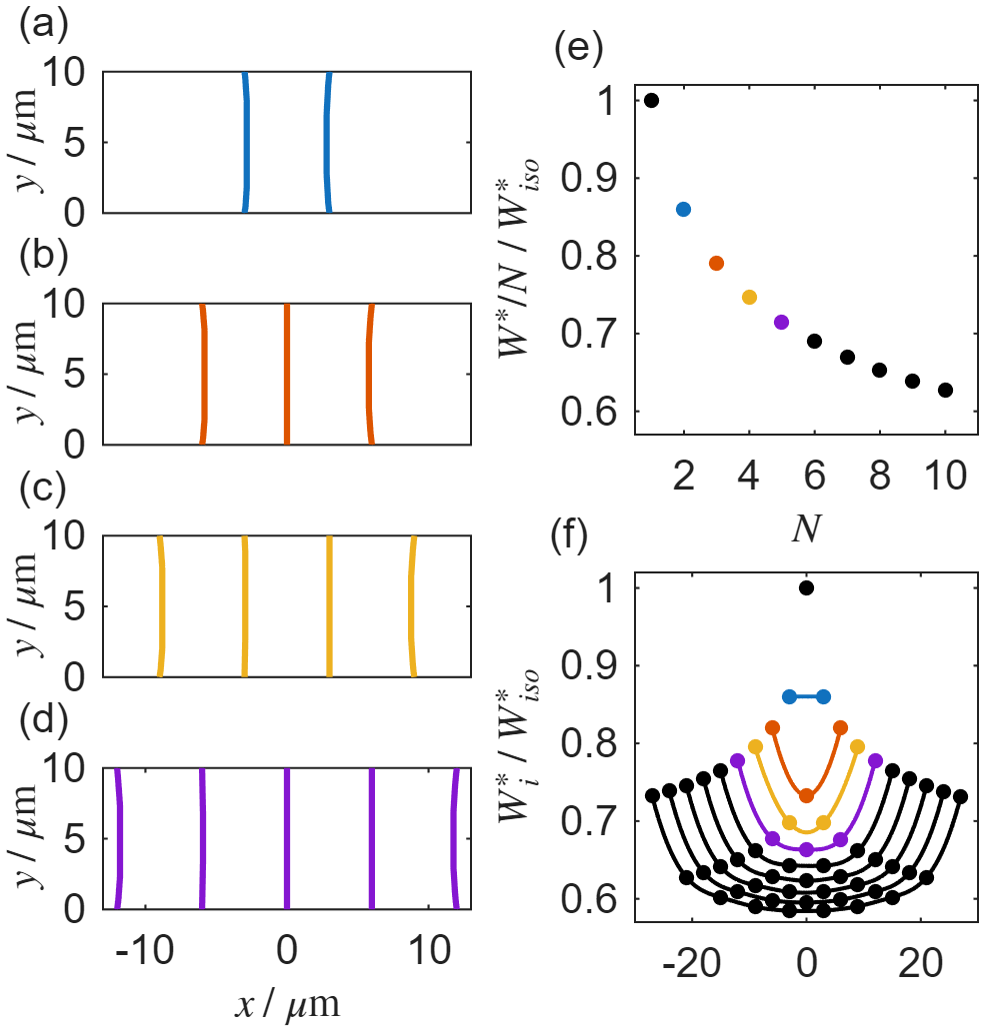}
    \caption{(a-d) Numerical prediction of optimal protocols for two, three, four and five traps that are shifted by 10~\unit{\micro\meter} in $y$ direction and spacing of 6~\unit{\micro\meter} on the $x$ axis. Viscosity and particle size are chosen as mentioned before in Fig.~\ref{fig:expres2d}. The qualitative strategies of approaching traps remains true for all particle numbers. The average optimal work per particle $W$ normalized to that of an isolated particle $W^*_\mathrm{iso}$ decreases with particle number as shown in (e) for up to 10 particles. The amount of work individual traps contribute depending on the starting position $\lambda_{x}^0$ is shown in (f). Contributions from one group are connected with a spline as a guide to the eye. For any groups the traps in the center require the least amount of work while traps at the edge require the most. Interestingly, in larger groups also the amount of work required for the driving at the edge reduces with group size, which confirms the long range nature of hydrodynamic interactions.}
    \label{fig:simNP}
\end{figure}
\indent When comparing the experimental data with numerical simulations (solid lines in Fig.~\ref{fig:expres2d}), we find excellent agreement. The simulations were fitted to each dataset using the solvent viscosity as the only adjustable parameter ($\eta_{\rightrightarrows}=6.0~\unit{\milli\pascal\second}$, $\eta_{\leftrightarrows}=6.2~\unit{\milli\pascal\second}$, and $\eta_{\rightarrow}=6.9~\unit{\milli\pascal\second}$), corresponding to a variation of about $8\%$. This small variation is attributed to weak optical absorption of the trapping beams by the silica particles, which slightly increases the local solvent temperature and thereby lowers the viscosity \cite{cheng2008formula,volk2018density}. Because this effect is stronger for two nearby particles than for a single particle, it naturally explains the minor differences between the datasets. In all cases, the extracted viscosities remain close to the literature value for a $1{:}1$ water--glycerol mixture ($\eta=6.9~\unit{\milli\pascal\second}$) \cite{cheng2008formula,volk2018density}. Remaining small deviations between experiment and simulation likely arise from minor experimental imperfections not included in the model, such as unavoidable particle polydispersity (standard deviation $0.06~\unit{\micro\meter}$) and spatial variations of the trap stiffness of about $5\%$.\\
\indent Having established how hydrodynamic interactions shape optimal protocols for two traps, we now extend our results to larger systems, where genuine collective behavior emerges. Each particle both dissipates energy into the fluid and modifies the dissipation experienced by all others, leading to a collective redistribution of viscous losses that favors coordinated motion. In Fig.~\ref{fig:simNP}(a-d), we show numerical predictions for optimal protocols involving up to five traps that are transported between prescribed initial and final configurations. In all cases, the traps share the same net displacement along the driving direction, while their initial and final positions differ only in their horizontal separations (co-propagating configuration). Again, each particle is driven by its own independently controlled optical trap, and the time-dependent protocols between the prescribed boundaries are determined by optimization. We assume the hydrodynamic interaction to be pairwise additive, which we confirmed in an control experiment involving three driven particles, see App.~\ref{sec:HImany} for details.\\
\indent The qualitative features found for two particles persist in larger assemblies composed of $N$ particles. Similar to the co-propagating case discussed above, the optimal protocols transiently reduce the lateral particle separations. This behavior reflects the long-ranged nature of hydrodynamic interactions, as each particle energetically benefits not only from its nearest neighbors but from the collective configuration of the entire assembly. This effect is quantified in Fig.~\ref{fig:simNP}(e), where we plot the energetic cost per particle, $W^*/N$, normalized by that of an isolated particle, $W^*_\text{iso}$, as a function of $N$. Already at $N=10$, the energetic cost is reduced by more than one third, highlighting particle interactions as a resource for reducing energetic expenditure. Additional insight is obtained by analyzing the work required to translate individual particles as a function of their position within the group. For all considered system sizes, particles located near the center of the assembly require the least work. Remarkably, even particles at the edges of the group are transported more efficiently when embedded in a larger collective than when moved in isolation.\\

\section{Outlook and Discussion}
We investigated theoretically and experimentally optimal finite-time control protocols for many-particle systems in which each particle is transported by an individually controlled harmonic trap. In the presence of interactions, particle dynamics become strongly coupled, substantially increasing the complexity of the control problem. Examination of the structure of the cost functional in the low-noise limit suggests a general qualitative distinction between conservative and non-conservative interactions.\\
\indent For conservative interactions, optimal control is universally characterized by straight-line particle trajectories with constant velocity. This reflects the fact that conservative forces contribute only boundary terms to the work functional and therefore do not affect the bulk dissipation. As a concrete illustration, we derived explicit optimal protocols for two particles coupled by a Hookean spring in the noise-free limit.\\
\indent This characteristic linearity breaks down for non-conservative interactions, as demonstrated experimentally using hydrodynamically coupled colloidal particles confined by optical tweezers. Numerical predictions and experiments for co- and counter-propagating traps show that optimal protocols generally involve curved trajectories, which either enhance or suppress interparticle coupling, depending on the particle configuration. While our experiments involve only two particles, the underlying mechanism naturally extends to larger assemblies: hydrodynamic interactions enable momentum transfer through the surrounding fluid, allowing forces applied to one particle to redistribute across the entire ensemble. In this way, hydrodynamic coupling can favor cooperative motion over independent translation and thereby reduce the energetic cost of transport. These qualitative features therefore persist in larger systems and highlight the collective nature of hydrodynamically assisted transport~\cite{weihs1973hydromechanics,gazzola2016learning,filella2018model}.\\
\indent Taken together, our results show that interparticle interactions are not merely a complication but can act as a resource for optimal control in many-particle systems. Depending on configuration and interaction type, interactions can substantially reduce energetic cost, as demonstrated here for co-propagating particles with hydrodynamic coupling. In the present experiments, hydrodynamic interactions were described within a pairwise additive approximation, which provides an accurate representation under the dilute conditions and moderate velocities considered in our experiments. In more general situations, however, they may become non-additive, particularly at smaller particle separations or higher velocities where higher-order hydrodynamic couplings become important~\cite{brady1988stokesion}. Similar non-pairwise effects arise for other field-mediated interactions, such as diffusiophoretic chemical gradients, thermophoretic temperature fields, electrokinetic flows, or optical binding forces between particles~\cite{saha2014clusters,michelin2014phoretic,sukhov2017non}. Incorporating such non-additive effects would introduce additional complexity into the control problem but may also open new possibilities for optimal control in many-body systems.\\ 
\indent Looking ahead, the present framework naturally invites extensions to partially actuated systems, environments with memory, and situations in which fluctuations cannot be neglected but must be actively exploited. At a broader level, the fluid-mediated momentum redistribution identified here is also related to the energetic advantages of collective locomotion observed in much larger systems, such as fish schools or bird flocks~\cite{weihs1973hydromechanics,gazzola2016learning,filella2018model}. More generally, these results point toward principled strategies for energy-efficient control of interacting microscopic systems, with potential relevance for soft microrobotics, active matter, and driven many-body systems.

\begin{acknowledgments}
We thank Joost de Graaf for insightful discussions on hydrodynamic interactions. We thank Peter Sollich for fruitful discussions on the influence of noise.  C.B. acknowledges funding by the European Research Council through the Advanced Grant BRONEB (No. 101141477).  S.L. acknowledges funding by UK Research and Innovation (UKRI) under the UK government’s Horizon Europe funding Guarantee (Grant No. EP/X031926/1) and from Corpus Christi College, Cambridge.
\end{acknowledgments}

\subsection*{Author contributions}
L.S. and S.L. performed analytical calculations. C.B and S.M. designed the experiment. S.M. performed the experiment and evaluated the data S.M., which was discussed together with C.B. and S.L.. Numerical calculations were performed by L.S. (spring model) and S.M. (hydrodynamic model). All authors wrote the manuscript.

\subsection*{Competing Interests}
The authors declare no competing interests.

\subsection*{Data availability}
The data that support the findings of this article are openly available at \cite{monter_2026_19046622}.

\appendix

\section{Experimental Methods}
\label{sec:expmeth}

\subsection{Optical trap setup}
\label{sec:setupNcali}

We use a 532\,\unit{\nano\meter} laser (Coherent, Verdi V2) as light source for the optical tweezers setup. 
The laser beam can be intensity-modulated and angle deflected in two axes with an acousto-optic deflector (AOD, AA opto-electronics DTSXY-400). 
The deflected beam is translated to the back aperture of the microscope objective (Olympus Apochromat MPLAPON-Oil 100x NA=1.45) with a telescope made of 2 inch lenses to account for the possible deflection angels of up to $\pm40~\unit{\milli\radian}$.
The microscope objective is used for tweezing as well as for imaging.
The laser power is adjusted so that it can reach a maximum of 160~\unit{\milli\watt} just in front of the back aperture of the objective when the AOD is run at the highest setting used in experiments.
For recording microscopy videos, a digital camera (Basler ace 2 a2A3840-45umPRO) is used. 
Videos are acquired at a frame rate of 100~\unit{\hertz} with a region of interest tailored to the requirements of the current experiment.
The sample temperature is controlled by resistive heating (Okolab H401-T-Penny) of the sample stage and the microscope objective. 
The temperature was kept at $25\pm0.05~\unit{\celsius}$ throughout all experiments.\\
\indent This setup makes it possible to shift the laser focus in the imaging plane of the objective by up to $\pm 19\,\unit{\micro\meter}$ in each dimension. 
Further, multiple optical traps can be realized through a time sharing approach: 
When the laser position is changed faster than the relaxation time of the particle in the optical trap, the light intensity appears to be a superposition of individual locations \cite{jones2015optical,gieseler2021optical}.
Thereby if one scans over multiple positions fast enough multiple optical traps can be created with just one beam.
We update the position at a rate of 10~\unit{\kilo\hertz} which is sufficiently fast compared to the relaxation time, which is on the order of ten of milliseconds.
To avoid artifacts from the AOD, a zero intensity gap is put in between two positions in the control signal of the AOD, i.e., a control loop looks like (position 1, zero, position 2, zero, position 1,...).\\
\indent Converting between the camera field‑of‑view positions and the frequencies needed to control the AOD requires calibration measurements.
In such a calibration measurement, the particles are held statically at positions for 120~\unit{second}, covering a grid over the whole field of view. 
The measured average particle position is then set equal to the position of the laser focus at the given AOD frequency. 
To reach positions that are not tested in the calibration measurement, a linear fit can be used to generate a function that maps frequencies to positions or vice versa.
The acquired data can also be used to compensate for differences in trapping stiffness within the focal plane by adjusting the AOD amplitude $A$.
This is done by measuring the trap stiffness $\kappa$ in $x$- and $y$-direction at each position of the calibration measurement. 
For every frequency pair $(f_x,f_y)$ a relative trap strength 
\begin{equation}
    \kappa_\mathrm{eff}(f_x,f_y) = \frac{(\kappa_x/\la \kappa_x\ra + \kappa_y/\la \kappa_y\ra}{2}
\end{equation}
is calculated.
Here $\la\kappa_i\ra$ is the average of all $\kappa_i$ measured in either $x$ or $y$ dimension.
Then a two-dimensional polynomial of order three is fitted to $\kappa_\mathrm{eff}(f_x,f_y)$.
This fit $F_\kappa(f_x,f_y)$ can then be used as a calibration factor to adjust the applied AOD amplitude $A_\mathrm{apply}$ from an initial set value $A_\mathrm{set}$
\begin{equation}
    A_\mathrm{apply} = \frac{A_\mathrm{set}}{0.5 + 0.5\cdot F_\kappa(f_x,f_y)}.
\end{equation}
With this method, the variation of trap stiffness in the sample plane can be reduced from $\pm20~\%$ to $\pm5~\%$.
In experiments with two particles, the calibration functions for the two particles are determined separately.
This is possible since the field of view can be divided into a separate region for each particle.
The division halves the required time of a calibration measurement.
Additionally increases the quality of the calibration as the difference in particle size appears to cause problems when applying a calibration function measured with one particle to another.

\begin{figure}
    \centering
    \includegraphics[scale = 1]{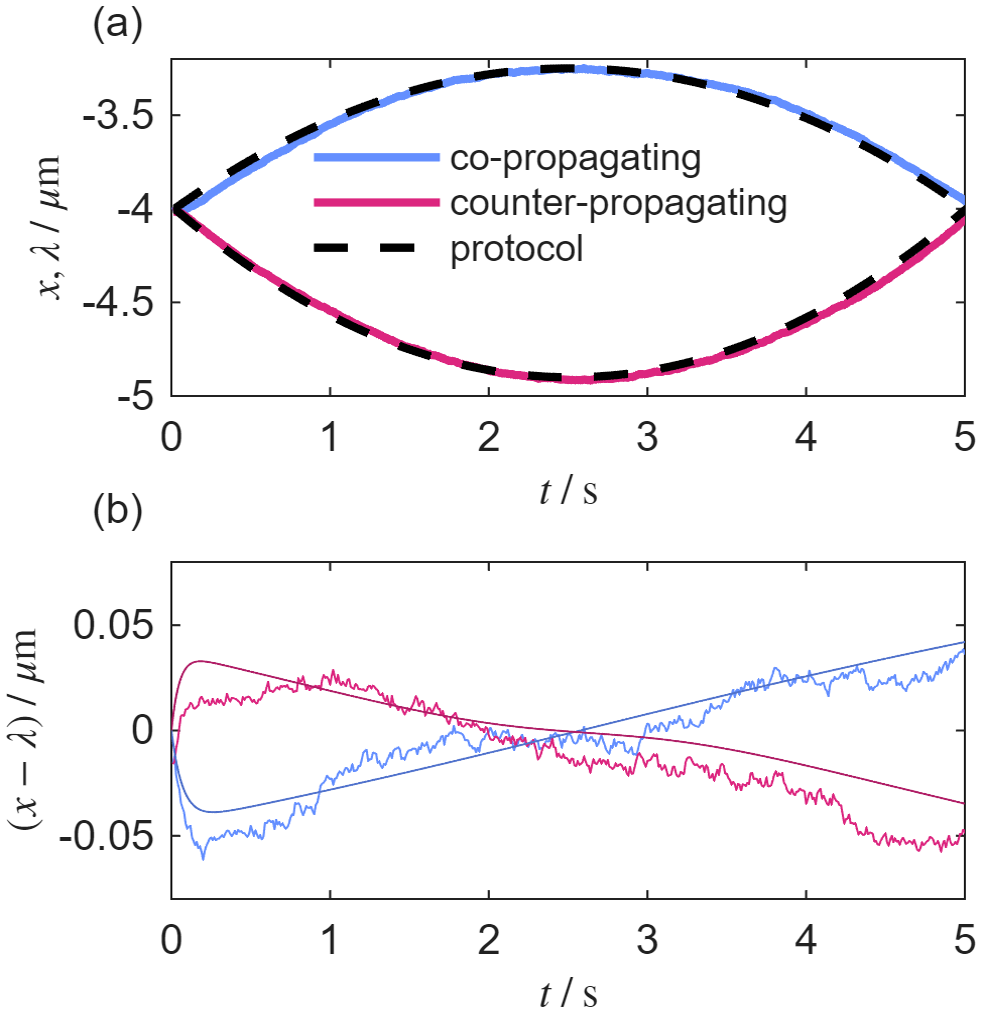}
    \caption{(a) Averaged particle trajectories in the $x$-dimension over time for $\lambda_p = 0.75~\unit{\micro\meter}$ for the co-propagating and $\lambda_p = -0.9~\unit{\micro\meter}$ for the counter-propagating configuration. Since protocol time is larger than the particle relaxation time in the trap, the particles on average closely follow the protocol shown as dashed black lines. (b) Difference between average particle position and trap center along the $x$-coordinate. Smooth and noisy lines correspond to simulations and averaged experimental data. The non-zero value for $t=0$ in the experimental data is due to remaining calibration errors and are close to the detection limit.}
    \label{fig:meantraj}
\end{figure}

\subsection{Sample preparation}
\label{sec:sample}
Sample solutions are prepared by dispersing spherical micro-particles (diameter $2.73\pm0.12~\unit{\micro\meter},\;\mathrm{SiO}_2$, microParticles GmbH) in a water-glycerol mixture with volume ratio of 1:1.
Sample cells are prepared by filling glass capillaries with an inner diameter of 100~\unit{\micro\meter} and a width of 1~\unit{\milli\meter} (CM Scientific) with the sample solution. 
The capillaries are closed with a combination of wax and epoxy resin. 
After filling the capillary, the samples are equilibrated in the measurement setup for at least 60 minutes.

\subsection{Measurement Procedures}
\label{sec:measproc}

After preparing the sample cell as described in App.~\ref{sec:sample}, two (or one) particles are caught in optical tweezers. 
The position of the objective along the optical axis is adjusted so that the particles have sufficient distance ($>30~\unit{\micro\meter}$) to the capillary walls.
Additionally, it is adjusted so that the trap stiffness for the two particles is roughly equal for both particles when they are held in opposite ends of the field of view used in the experiment.
Then a calibration measurement is started and the calibration functions are derived as explained in App.~\ref{sec:setupNcali}.
All protocols consist of moving the trap centers $\lambda_i$ according to a specific protocol.
A single experiment always consists of the forward and backward realization of the relevant protocol.
This prevents a continuous motion through the sample which would not be possible due to the limited range of the AOD and field of view of the camera.
Before, in between and after the forward and backward realizations, the traps are held in a static position for 3~\unit{second} to make sure to always start from an equilibrium state.
Single experiments like this were then repeated at least 150 times to compute meaningful averages.
Averaging over the back and forward versions of a protocol also allows us to average out variations of the trap stiffness that are not directly measurable.\\\indent
For measured trajectories, we can compute stochastic work according to \cite{nonequilibrium1998crooks}
\begin{equation}
    W = \sum_{n=0}^{N_\mathrm{t} -1}V\left\{\f r\left(n\Delta\right),\f \lambda\left[\left(n+1\right)\Delta\right]\right\} - V\left[\f r\left(n\Delta\right),\f \lambda\left(n\Delta\right)\right],\label{eq:Wcrooks}
\end{equation}
where the time is partitioned into discrete intervals of size $\Delta$ with a total of $N_\mathrm{t} = t_\mathrm{f}/\Delta$ steps.

\section{Hydrodynamic Interaction}
\label{sec:HI}

For a system of $N$ particles suspended in a fluid, it is insufficient to describe each particle with its own Langevin equation, since particles are coupled through the fluid.
A movement of one particle displaces the surrounding fluid, which induces flow fields that exert forces on other particles nearby.
To represent this coupling, the constant mobility in the single-particle equation is replaced by a position dependent mobility tensor $\HM$.
The diagonal elements of $\HM$ give the mobility of isolated particles, while the off-diagonal elements describe pairwise interactions.
These interactions depend on the fluid viscosity and, crucially, on the relative positions of the particles.
$\HM$ can be written as a $N \times N$ block matrix with entries $\f\mu_{ij}$, representing the interaction between particles $i$ and $j$.
Recall that $d$ is the spatial dimension, so that each $\f\mu_{ij}$ itself is a $d\times d$ tensor.
The complete mobility tensor therefore is of rank $M \times M$ where $M=N\,d$.
We compute $\f\mu_{ij}$ using the Rotne–Prager–Yamakawa approximation (RPY) \cite{rotne1969variational,dhont1996introduction}, obtained by integrating the Stokes flow generated by a rigid sphere over its finite volume
\begin{align}
	\textbf{H} =&
        \begin{pmatrix} \f \mu_{11} \dots \f  \mu_{1N} \\
		      \vdots\quad\ddots\quad\vdots \\
		      \f \mu_{N1}\dots \f \mu_{NN} 
        \end{pmatrix}\,, &&\label{eq:H} \\ 
	\f \mu_{ij} =&\mu_0 \left[ \frac{3 a}{4 \mathrm{r}_{ij}} \left(\textbf{I} + \frac{\f r_{ij}\otimes \f r_{ij}}{\mathrm{r}_{ij}^2}\right)\right.+ &&  \\ 
	&\left. \frac{1}{2}\left(\frac{a}{\mathrm{r}_{ij}}\right)^3  \left(\textbf{I} - 3\frac{\f r_{ij}\otimes \f r_{ij}} {\mathrm{r}_{ij}^2}\right) \right]\,, &\;\mathrm{for} & \;i \neq j   \label{eq:Hoffdiag}\,, \\ 
	\f\mu_{ij}  =& \mu_0 \textbf{I}\,,& \; \mathrm{for} &\; i = j \,.\label{eq:Hdiag}
\end{align}
Here, $\fr_{ij}$ is the vector that connects the position of the particle $i$ and $j$, $\mathrm{r}_{ij}$ is the distance between the particles, 
$a$ is the radius of the particle, and $\mu_0 = (6\pi\eta a)^{-1}$ the mobility of a single particle with the fluid viscosity $\eta$. The outer product is denoted $\otimes$ and the unity matrix of matching size is written as $\textbf{I}$.
The mobility tensor can be simplified according to its definition.  
For diagonal blocks, i.e., with $i = j$, $\f\mu_{ii}$ is a diagonal matrix.
Where the diagonal entries are the bulk mobility of a single particle.
Further, $\f \mu_{ij}$ is symmetric in particle and dimension exchange, i.e., $\mu^{xy}_{ij} = \mu^{xy}_{ji}$ and $\mu^{xy}_{ij} = \mu^{yx}_{ij}$.
Here, the superscript shows the coupled dimensions and the subscript the coupled particles for an individual entry of $\f\mu_{ij}$.
For the special case of two particles in two spatial dimensions, we can also simplify $\f\mu_{ij}$ for $i \neq j$ by substitution: $\mu^\parallel = \mu^{xx}_{ij}$ and $\mu^\perp = \mu^{yx}_{ij}$.
This leads to 3 non-zero entries in the mobility tensor
\begin{equation}
    \HM =
	\begin{pmatrix} 
		\mu_0             &0                   &\mu^{\parallel}   &\mu^{\perp}\\
		0               &\mu_0                 &\mu^{\perp}   &\mu^{\parallel}\\ 
		\mu^{\parallel}  &\mu^{\perp}       &\mu_0             &0\\ 
		\mu^{\perp}   &\mu^{\parallel}       &0               &\mu_0\\\end{pmatrix}.
\end{equation}
The bulk friction $\mu_0$ transforms a force in a direction on a particle into a velocity of that particle in the same direction. The contributions from $\mu^\parallel$ transforms a force on one particle into a velocity of the other particle in the same direction.
Moreover, $\mu^\perp$ transforms a force on a particle into an orthogonal velocity on another particle. 
Note that while $\mu_0$ is constant, $\mu^{\parallel}\; \text{and} \; \mu^{\perp}$ depend on the relative position of the two particles.

\subsection{Noise average}
\label{sec:HInoise}
When formulating the Langevin equation for an overdamped system of hydrodynamically coupled particles, special care needs to be taken in order to reproduce the correct physical behavior.
More precisely, an additional term including the divergence of the mobility tensor $\nabla\cdot\HM$ is necessary to reach the correct equilibrium distribution of the system.
The resulting equation then reads
\begin{equation}
    \dot{\f R} = -\HM \frac{\partial V_{\mathrm{trap}}}{\partial \f R}  + k_\mathrm{B}\,T \;\nabla \cdot \textbf{H} + \sqrt{2k_\mathrm{B}\,T}\; \textbf{B} \f \zeta(t)\,.\label{eq:langevinHInoise}
\end{equation}
Here, $\textbf{B}\textbf{B}^T = \textbf{H}$ and $\f \zeta(t)$ is Gaussian white noise with zero mean $\langle\f \zeta(t)\rangle = \f0$ and $\langle \f\zeta_i(t)\f\zeta_j(s)^T\rangle = \delta_{ij}\textbf{I}\delta(t-s)$.
The calculation of the noise average of this equation is in general not straightforward due to the non-linear character of $\HM$.
However, we operate in a dilute system, which means that the minimal particle distances are still on the order of the particle diameter and the number of particles is low ($2\leq N\leq10$).
In this dilute regime, we find that the particle interactions are pairwise additive (see Sec.~\ref{sec:HImany} for details), which means that the divergence of the mobility tensor $\nabla\cdot\HM$ vanishes \cite{reichert2006hydrodynamic,wajnryb2004brownian}.
At the same time, the fluctuations of the particles in their confinement are small compared to the distance they travel during the control problem.
Under these constraints, we can use a simplified version of the noise average by assuming that \textbf{B} is linear on the small scale of fluctuations.
This makes it possible to drop the last two terms in the noise average, which leads to the averaged equation of motion reading
\begin{equation}
    \dot{\fr} = -\textbf{H} \frac{\partial V_{\mathrm{trap}}}{\partial \fr}\,. \label{eq:langevinHIaverage}
\end{equation}
This reads the same as the athermal equation of motion for hydrodynamically coupled overdamped spheres.
We can test this assumption by comparing simulations with and without noise. 
We choose parameters that are similar to the problems under investigation.
When comparing the average of many simulations of the full stochastic equation of motion [see Eq.~\eqref{eq:langevinHInoise}] to results from the simplified form [see Eq.~\eqref{eq:langevinHIaverage}] without noise, we observe no systematic difference.
Further, the distribution of noisy trajectories is gaussian.
This allows us to compare numerical results to averaged experimental data which is subject to thermal fluctuations.

\subsection{Many-Body effects}\label{sec:HImany}

\begin{figure*}
    \includegraphics[width = 0.9\textwidth]{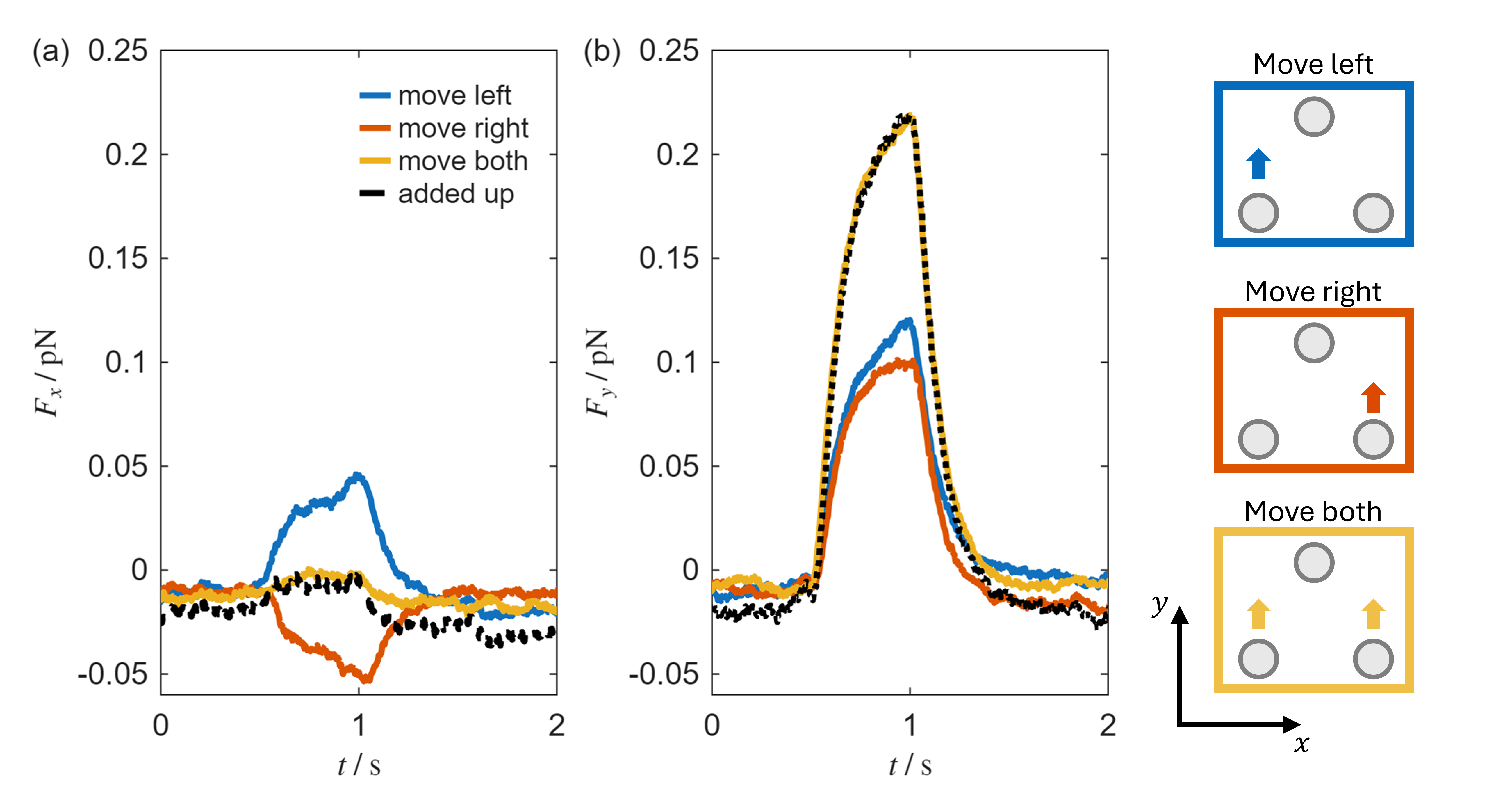}
    \caption{We confine three particles in a triangular configuration using optical tweezers, with positions $r_\mathrm{left}/\unit{\micro\meter} = (-5.2,-3.7)$, $r_\mathrm{right}/\unit{\micro\meter} = (5.5,-3.7)$, and $r_\mathrm{center}/\unit{\micro\meter} = (0.3,3.8)$ as shown in the sketches on the right. We then displace the right particle, the left particle, or both particles simultaneously, by $\Delta r_\mathrm{left}/\unit{\micro\meter} = (0,3.0)$ and $\Delta r_\mathrm{right}/\unit{\micro\meter} = (0.0,3.7)$, while measuring the resulting force exerted on the central particle. The force curve obtained when both particles are displaced can be quantitatively reconstructed as the sum of the individual force curves measured when only one particle is moved at a time. As shown for the force components in $x$-dimension (a) and $y$-dimension (b). The small but finite residual force observed in the absence of external perturbations originates from an asymmetry (skewness) in the trapping potential, which is not captured by the harmonic approximation employed to determine the trap stiffness.}
    \label{fig:manybody}
\end{figure*}

In the last part of our study, we numerically predict optimal protocols for more than two particles, and we assume that the hydrodynamic interactions are pairwise additive.
Numerical studies suggest that additional many-body effects become relevant only when particle separations are smaller than the radius of the particle \cite{torre2025hydrodynamic}.\\\indent 
We tested this in experiments with three particles.
As shown in Fig.~\ref{fig:manybody} we hold three particles in a triangular configuration with the optical tweezers [$r_\mathrm{left}/\unit{\micro\meter} = (-5.2,-3.7),\,r_\mathrm{right}/\unit{\micro\meter} = (5.5,-3.7)\,\text{and}\,r_\mathrm{center}/\unit{\micro\meter} = (0.3,3.8)$].
We then shift either the right, the left, or both particles upward by translating the corresponding trap over a distance of 3.7~\unit{\micro\meter} at constant velocity within one second.
During this, we measure the resulting force on the particle in the center (see Fig.~\ref{fig:manybody}).
We can reproduce the force curve measured when two particles are moved by adding the individual force curves resulting from the movement of either single particle.
This is a strong argument for the pairwise additive nature of hydrodynamic interactions in the regime in which we perform experiments.
The non zero force remaining when the system is not disturbed arises from a skewness in the potential that is not covered by the harmonic fit used to deduct the trap stiffness.
Note that since the absolute values of the forces acting on the moved particles do not matter, we did not perform the standard calibration experiments (see App.~\ref{sec:setupNcali}) for this test.

\section{Numerical Optimization}
\label{sec:numerics}

As cost functional for our optimization problem of shifting multiple harmonic potentials holding overdamped particles over a given distance in a finite time $\tf$ we choose the definition of work from stochastic thermodynamics [see Eq.~\eqref{Work}].
After partial integration and plugging in the noise averaged equation of motion for hydrodynamically coupled systems Eq.~\eqref{eq:langevinHIaverage} we can rewrite the cost functional as
\begin{align}
    W &= \int_0^{\tf} \dot{\f r}^T\HM^{-1}\dot{\fr} \;\dt\;+\;\Delta V \nonumber\\
    &= \int_0^{\tf} \left(\HM \frac{\partial V_{\mathrm{trap}}}{\partial \fr}\right)^T\frac{\partial V_{\mathrm{trap}}}{\partial \fr} \;\dt\;+\;\Delta V\,.
\end{align}
Here $\frac{\partial V_{\mathrm{trap}}}{\partial \fr} = k(\f\lambda-\fr)$ is the force  exerted on the particle by the confining potential and $\Delta V$ is the difference in trap potential between beginning and end of the control period.
From the cost functional, we can extract the running cost $L = \left(\HM \frac{\partial V_{\mathrm{trap}}}{\partial \fr}\right)^T\frac{\partial V_{\mathrm{trap}}}{\partial \fr}$ and use it to construct the optimal control Hamiltonian \cite{meinsma2023course,alvarado2025optimal}
\begin{equation}
    \mathcal{H} = \f p^T\dot{\fr} - L \label{eq:optconham}\,.
\end{equation}
Here $\f p$ is the so called costate.
In the next step, we derive the optimal control Hamiltonian equations
\begin{equation}
    \dot{\fr} = \frac{\partial \mathcal{H}}{\partial\f p} ,~~~\dot{\f p} = -\frac{\partial \mathcal{H}}{\partial \fr}, ~~~\text{and}~~~\f0 = \frac{\partial \mathcal{H}}{\partial \f\lambda}\,.\label{eq:HamEq}
\end{equation}

Rearranging the last equation in Eq.~\eqref{eq:HamEq} yields an expression for the optimal control $\f\lambda^*$ as a function of $\f r$ and $\f p$.
The analytical derivation up to this point is done with SymPy \cite{sympy}, which makes it easy to use the resulting expressions in the following numerical routine.
The solutions to the derived differential equations $\dot{\f r}$ and $\dot{\f p}$ also include the optimal solution to our control problem.
To determine the optimal solution, the cirrect initial conditions of $\fr$ and $\f p$ need to be found.
Since the protocol starts from equilibrium, we know the initial condition for $\fr$.
However, the initial condition for the costate ${\f p_0}$ is unknown.
To find ${\f p_0}$ we formulate a cost function that takes values of ${\f p_0}$ and then numerically solves ${\fr}$ and ${\f p}$ as an initial value problem.
We can integrate $\dot{\fr}$ and $\dot{\f p}$ numerically by including the calculation of $\f\lambda^*$ in the definition of the right hand side of the system of differential equations.
We perform the integration using the SciPy ivp\_solver function \cite{scipy} and choose RK45 as integration method \cite{dormand1980family,shampine1986some}, as it yields the most stable behavior.
From the resulting particle trajectories $\f r(t)$ and costate $\f p(t)$ we can derive $\f\lambda(t)$.
From $\fr(t)$ and $\f\lambda(t)$ we can calculate the required work according to Eq.~\eqref{eq:Wcrooks} \cite{nonequilibrium1998crooks} as the output of our cost function.
To incorporate jumps at the beginning and end of the protocol, we attach the known initial and final positions of $\f\lambda$ and repeat the first and last positions of $\fr$, assuming that the particle remains static during an instantaneous jump.
We then optimize this cost function using the SciPy minimize function \cite{scipy} using the Powell algorithm \cite{efficient1964powell,brent2013algorithms}.\\\indent
To test the validity of our method, we calculate optimal protocols for single particles, negligible interactions i.e. large particle distances ($>100a$) in the case of hydrodynamic interactions, and compare them to the known analytical results \cite{optimal2007schmiedl}.
Additionally, we can compare the numerical results for two particles coupled with a linear spring with zero rest length to our own analytical results.
In all cases, we find quantitative agreement up to the limit of numerical precision.

\subsection{Symmetry Simplification}

When formulating the optimal control Hamiltonian for two particles we can make use of the symmetry of the problems investigated in experiments.
In the case of two particles shifted side by side, i.e. the co-propagating configuration, the system has an axial symmetry, and the case where two particles are shifted against each other, i.e. counter-propagating configuration, the system has a point symmetry.
This makes it possible to half the number of differential equations that need to be solved, which significantly speeds up the numerical process. 
\paragraph{Co-propagating configuration:} In the co-propagating configuration the position vectors can be reduced to 
\begin{equation}
    \fr  =\begin{pmatrix}
        x \\
        y \\
        -x\\
        y \\
    \end{pmatrix}, ~~
    \f\lambda  =\begin{pmatrix}
       \lambda_x \\
        \lambda_y \\
        -\lambda_x\\
        \lambda_y \\
    \end{pmatrix}, ~~\text{and}~~
    \f p  =\begin{pmatrix}
        p_x \\
        p_y \\
        -p_x\\
        p_y \\
    \end{pmatrix}.
\end{equation}

\paragraph{Counter-propagating configuration:}
In the counter-propagating configuration  the position vectors can be reduced to 
\begin{equation}
    \fr  =\begin{pmatrix}
        x \\
        y \\
        -x\\
        -y \\
    \end{pmatrix}, ~~
    \f\lambda  =\begin{pmatrix}
       \lambda_x \\
        \lambda_y \\
        -\lambda_x\\
        -\lambda_y \\
    \end{pmatrix}, ~~\text{and}~~
    \f p  =\begin{pmatrix}
        p_x \\
        p_y \\
        -p_x\\
        -p_y \\
    \end{pmatrix}.
\end{equation}

\section{Optimal Control of Systems with Conservative Interactions at Zero Noise}\label{AppendixConservative}
Here, we provide the details for the result concerning temporal and geometric linearity (steady straight-line motion) of optimal particle trajectories in many-body systems with conservative interactions. We initially state the problem with thermal noise, to illustrate the consequences of the interplay of noise and nonlinear forces. We then take the athermal limit $T\to 0$, which allows us to find an analytical solution. 

We recall that we consider $N$ particles at positions $\fx_i\in\mathds{R}^d$, $i=1,\dots,N$, which are dragged by optical traps with stiffness $\kappa$ and trap centers $\fl_i\in\mathds{R}^d$. Since the traps can be moved according to a control protocol in all spatial directions, we have one control parameter $\fl_i$ per system degree of freedom $\fx_i$. We assume that the position of the $i$th particle evolves according to the underdamped Langevin equation
\begin{align}
    \rmd \fx_{i,s} &= \f{v}_{i,s}\rmd s\\
    m_i\rmd\f{v}_{i,s}&=-\f{\gamma}_i\f{v}_{i,s}\rmd s - \nabla_i V\rmd s + \sqrt{2k_\mathrm{B}T\f\gamma}\rmd \f W_{i,s}\nonumber
\end{align}
where $m_i$ denotes the colloidal mass, $\f\gamma_i$ are the symmetric positive definite friction matrices, $\nabla_i$ is the gradient with respect to the position of particle $i$, and the total potential energy is a differentiable function that can be decomposed into 
\begin{align}
     V(\{\fx_{i}\}, \{\fl_i(s)\}) = V_\mathrm{trap} + V_\mathrm{int} + V_\mathrm{ext},
\end{align}
with 
\begin{align}
    V_\mathrm{trap}(\{\fx_{i}\}, \{\fl_i\}) &= \frac{\kappa}{2}\sum_{i=1}^N\norm{\fx_i - \fl_i}^2,\\
    V_\mathrm{int}(\{\fx_{i}\}) &= \sum_{i=1}^N\sum_{j=i+1}^N V_\mathrm{int}^{ij}(\fx_i-\fx_j),\\
    V_\mathrm{ext}(\{\fx_{i,s}\})&= \sum_{i=1}^NV_\mathrm{ext}^i(\fx_i).
\end{align}
Here, $V_\mathrm{int}^{ij}=V_\mathrm{int}^{ji}$ accounts for the interaction between the $i$th and $j$th particle, which may differ from pair to pair. Further, $V_\mathrm{ext}^i$ accounts for possible external fields acting on each particle $i$. 

We denote the average position of the colloids by $\fr_i(s) = \langle \fx_{i,s}\rangle$, where the average is over the noise history. Using this, the trap position of the $i$th colloid can be written as
\begin{align}
    \fl_i(s) =& \frac{1}{\kappa}\big[m_i\ddot{\fr}_i(s) + \f\gamma_i\dot{\fr}_i(s) + \langle \nabla_i V_\mathrm{int}(\{\fx_i(s)\}) \rangle \big] \nonumber\\&+ \frac{1}{\kappa}\langle \nabla_i V_\mathrm{ext}(\{\fx_i(s)\})\rangle+ \fr_i(s).\label{LambdaConsApp}
\end{align}
The average work is therefore
\begin{align}
    W&=\kappa\int_0^{t_\mathrm{f}}\rmd s \sum_{i=1}^N\dot{\fl}_i^T\left[\fl_i-\fr_i\right]\nonumber\\
    =&\kappa\sum_{i=1}^N \fl_i^T\left[\fl_i-\fr_i\right]\Big|_0^{t_\mathrm{f}} -\kappa\int_0^{t_\mathrm{f}}\rmd s \sum_{i=1}^N\fl_i^T\left[\dot{\fl}_i-\dot{\fr}_i\right]\nonumber\\
    =&\kappa\sum_{i=1}^N \fl_i^T\left[\frac{1}{2}\fl_i-\fr_i\right]\Big|_0^{t_\mathrm{f}}\nonumber\\&+\kappa\int_0^{t_\mathrm{f}}\rmd s \sum_{i=1}^N\left\{\frac{1}{\kappa}\left[ m_i\ddot{\fr}_i + \f\gamma_i\dot{\fr}_i + \langle \nabla_i V_\mathrm{int}(\{\fx_i\})\right.\right.\nonumber\\ &+\left.\left.\nabla_i V_\mathrm{ext}(\{\fx_i\})\rangle \right] + \fr_i\right\}^T\dot{\fr}_i\nonumber\\
    =&\frac{\kappa}{2}\sum_{i=1}^N \norm{\fl_i-\fr_i}^2\Big|_0^{t_\mathrm{f}} +\frac{1}{2}\sum_{i=1}^Nm_i\dot{\fr}_i^T\dot{\fr}_i\Big|_0^{t_\mathrm{f}} \label{WorkNonlinearAverage}\\&+\int_0^{t_\mathrm{f}}\rmd s \sum_{i=1}^N\left[\f\gamma_i\dot{\fr}_i + \langle\nabla_i V_\mathrm{int}(\{\fx_i\})+\nabla_i V_\mathrm{ext}(\{\fx_i\}) \rangle\right]^T\dot{\fr}_i\nonumber
\end{align}
In the presence of nonlinear gradients of the potentials, we cannot readily evaluate the averages, making further analytical treatment very difficult. We discuss this case separately in Appendix \ref{noiseCorrection}. In the following, we restrict our discussion to cases where either (i) all forces are linear, or (ii) the system is deterministic ($T\to 0$), so that we can simplify the last expression using 
$\langle V_*(\fx_i)\rangle = V_*(\fr_i)$. In these cases, we can use
the chain rule to write the temporal derivatives of the potentials as
\begin{align}
    \frac{\rmd}{\rmd s}V_\mathrm{int}(\{\fr_i\})&= \sum_{i=1}^N \left[\nabla_i V_\mathrm{int}(\{\fr_i\})\right]^T\dot{\fr}_i\,,\\
    \frac{\rmd}{\rmd s}V_\mathrm{ext}(\{\fr_i\})&= \sum_{i=1}^N \left[\nabla_i V_\mathrm{ext}(\{\fr_i\})\right]^T\dot{\fr}_i.
\end{align}
Using these ingredients, the integral in Eq.~\eqref{WorkNonlinearAverage} in the fully linear (i) or athermal (ii) case can readily be expressed as
\begin{align}
    \!\!&\int_0^t\rmd s \sum_{i=1}^N\left[\f\gamma_i\dot{\fr}_i + \nabla_i V_\mathrm{int}(\{\fr_i\})+\nabla_i V_\mathrm{ext}(\{\fr_i\})\right]^T\dot{\fr}_i \nonumber\\
    \!\!=&V_\mathrm{int}(\{\fr_i\})\Big|_0^t+ V_\mathrm{ext}(\{\fr_i\})\Big|_0^t+\int_0^t\rmd s\sum_{i=1}^N \dot{\fr}_i^T\f\gamma_i\dot{\fr}_i\,.
\end{align}

Identifying the difference between initial and final potential and kinetic energy as
\begin{align}
    \Delta V \equiv& V_\mathrm{trap}(\{\fr_i(s)\}, \{\fl_i(s)\})+V_\mathrm{int}(\{\fr_i(s)\})\nonumber\\&+V_\mathrm{ext}(\{\fr_i(s)\})\Big|_{s=0}^{s={t_\mathrm{f}}},\nonumber\\
    \Delta E_\mathrm{kin} \equiv& \frac{1}{2}\sum_{i=1}^Nm_i\underbrace{\dot{\fr}_i(s)^T\dot{\fr}_i(s)}_{=\norm{\dot{\fr}_i(s)}^2}\Big|_{s=0}^{s={t_\mathrm{f}}},
\end{align}
allows us to write the work in a compact way
\begin{align}
    W = \Delta V + \Delta E_\mathrm{kin} + \int_0^{t_\mathrm{f}}\rmd s \sum_{i=1}^N\dot{\fr}_i^T\f\gamma_i\dot{\fr}_i.\label{ConservativeWork}
\end{align}
To find the optimum, we now have to minimize the temporal bulk contribution (integral term). This can be done in a straightforward way using the Euler-Lagrange equations of the integrand, which admits the solution $\ddot{\fr}_i^*(s)=\f0$ for all $i=1,\dots,N$. This implies that the optimal particle trajectories are given by $\fr_i^*(s) = \f c_i s + \fr_i(0)$ with some time-independent vectors $\f c_i\in\mathds{R}^d$ found by minimizing Eq.~\eqref{ConservativeWork} on the space of solutions $\fr_i^*(s) = \f c_i s + \fr_i(0)$. 
Thus, the optimal dynamics of the particles is a straight-line motion (``geometric linearity'') with constant velocity (``temporal linearity''). This is a general result for conservative, pairwise interactions, given that (i) either all forces are linear, or (ii) the noise is zero, in which case arbitrary nonlinearities in the interaction and external forces are allowed.

We emphasize that the temporal and geometric linearity of the particle trajectories does not imply linear trap protocols. This distinction is evident from Eq.~\eqref{LambdaConsApp}, where nonlinearities in both time and space can arise through conservative interaction and external fields. In the following section, we demonstrate that even purely harmonic interaction potentials between two colloids are sufficient to generate such nonlinear trap protocols.

Note that, although we assumed that the friction and mass of each particle is the same in the calculation above, the result that optimal particle paths are steady straight lines, easily generalizes to anisotropic mass and friction, e.g., if the colloids have different shape and sizes. Moreover, the result is also valid in the overdamped limit. While the optimal colloid trajectory remain linear regardless of the presence or absence of inertial effects, the optimal trap protocols are in fact greatly affected by the overdamped limit, see Ref.~\cite{gomezmarin2008underdamped}.

The physical interpretation of Eq.~\eqref{ConservativeWork} aligns with the general intuition for conservative systems: All direct potential dependence of the work is encapsulated in the boundary terms accounting for differences in potential and kinetic energy between the beginning and end of the process. The remaining bulk term, which depends on the actual path taken as the particle moves through the viscous medium, accounts solely for frictional dissipation. Consequently, straight-line motion—which minimizes the path length in flat configuration space—emerges as the universal optimal solution for the particle trajectory, irrespective of the specific form of the potential.

However, stepping beyond the fully actuated control setting considered here, situations with fewer control parameters than physical degrees of freedom need not exhibit constant-velocity optimal motion, 
even though geometric linearity of the trajectories may persist due to underlying symmetries of the system. A prominent example arises when the surrounding medium contains hidden degrees of freedom that participate in the dynamics but are not directly controllable, as in viscoelastic fluids. In such cases, temporal linearity generically breaks down, even for a single particle \cite{loos2024universal}.


\renewcommand{\l}{\lambda}
\newcommand{\px}{{\partial_{x_0}}}
\newcommand{\pxt}{\partial_{x_0}^2}
\newcommand{\pr}{{\partial_{r_0}}}
\newcommand{\prt}{\partial_{r_0}^2}
\newcommand{\Fx}{F(x_0)}
\newcommand{\Fr}{F(r_0)}

\section{Optimal Control of Systems with Conservative Interactions---Noise Corrections}\label{noiseCorrection}
As shown in Appendix~\ref{AppendixConservative}, the optimal colloid trajectories for $T=0$ in a conservative system are linear in both space and time. We will now use a perturbative approach to investigate what occurs in the $0<T\ll1$ case. For this, we use the noise perturbation approach found in e.g. Ref.~\cite{gardiner2004handbook} on a one-dimensional system with a single overdamped colloid. This is done for notational simplicity, and it should be clear that extensions to higher dimensions and/or multiple colloids and/or underdamped colloid(s) is straightforward (although with some technical details when introducing matrices and tensors). 

The generic overdamped Langevin equation with a moving optical trap reads
\begin{align}
    \rmd x_s = -\kappa (x_s-\l(s))\rmd s + F(x_s)\rmd s + \sigma\rmd W_s,\label{App1dODlangevin}
\end{align}
where $F(x) = -\partial_x V(x)$ is the gradient of some generally non-quadratic potential and $\sigma = \sqrt{2k_\mathrm{B}T}$ is the noise amplitude. Since we are interested in a low noise approximation, we will expand Eq.~\eqref{App1dODlangevin} in $\sigma\propto T^{-1/2}$. Specifically, we assume that the following series expansion can be made
\begin{align}
    x_{s} = x_{0,s} + \sigma x_{1,s} + \sigma^2x_{2,s}+\dots.\label{AppNoiseExp}
\end{align}
As discussed in Ref.~\cite{gardiner2004handbook}, this expansion is not always convergent depending on the potential $V(x)$, hence, we implicitly assume that it is convergent. With the noise expansion Eq.~\eqref{AppNoiseExp}, the gradient can be expanded as
\begin{align}
    F(x)& = F(x_0) +\sigma x_1\px F(x_0)\nonumber \\&+ \frac{\sigma^2}{2}\left\{2x_2\px F(x_0) + x_1^2[\pxt\Fx]\right\}+\dots  
\end{align}
Furthermore, we will also assume that the optimal protocol $\l^*$ can be expanded in $\sigma$
\begin{align}
    \l^* = \l_0^* + \sigma\l_1^* + \sigma^2\l_2^*+\dots
\end{align}
This ansatz is justified, if the zero-noise limit is known, and the system is sufficiently nicely behaved, so that a small noise amplitude only leads to a small correction in the optimal protocol. We will demonstrate below for the canonical example system studied in Ref.~\cite{optimal2007schmiedl}, that this approach leads to correct results. 

To calculate the leading order correction of the work in $\sigma$, we expand Eq.~\eqref{App1dODlangevin} and match orders of $\sigma$ \cite{gardiner2004handbook} (we suppress time-indices and the optimality superscript which should be clear from context)
\begin{align}
    \!\!\rmd x_0 &= -\kappa (x_0-\l_0)\rmd s + \Fx\rmd s\,,\\
    \!\!\rmd x_1 &= -\kappa(x_1 - \l_1)\rmd s + x_1\px\Fx\rmd s + \rmd W\label{x1Expansion}\,,\\
    \!\!\rmd x_2&=-\kappa(x_2-\l_2)\rmd s + \frac{1}{2}\left[2x_2\px\Fx + x_1^2\pxt\Fx\right]\nonumber\,,\\
    &\vdots\nonumber
\end{align}
Now we perform the noise averages, denoting $r=\langle x\rangle$ and $r_j=\langle x_j\rangle$. At order $\sigma^0$, 
we can perform the noise average in a straightforward manner, and obtain a deterministic equation for $r_0=\langle x_0\rangle = x_0$. At order $\sigma^1$, the dynamics is an Ornstein-Uhlenbeck process with time-dependent coefficients, where the nonlinearities only depend on the zeroth order solution. 
Therefore, the system is linear in the stochastic variable $x_1$ and the noise average can again be taken in a straightforward manner. 
Only for orders $\sigma^n$, $n\geq 2$, do fluctuations and nonlinearities in the forces interfere, so that the dynamics involves nontrivial correlation functions:

\begin{align}
    \dot r_0 &= -\kappa (r_0-\l_0) + \Fr\label{AppNoiseExr0}\,,\\
    \dot r_1 &= -\kappa(r_1 - \l_1) + r_1\pr\Fr\label{AppNoiseExr1}\,,\\
    \dot r_2 &= -\kappa(r_2 - \l_2) + r_2\pr\Fr + \frac{1}{2}\langle x_1^2\rangle\prt\Fr\label{AppNoiseExr2}\,,\\
    &~\,\vdots
\end{align}

\subsection{Expansion of Work to First Order}
We now insert the expansion \eqref{AppNoiseExr0}-\eqref{AppNoiseExr1} into the work functional and neglect all corrections of order $\mathcal{O}(\sigma^2)$. To keep notation compact, equalities are to be understood as ``\textit{equal up to order $\mathcal{O}(\sigma^2)$}'' in the following. We find
\begin{align}
    W =& \left\langle \kappa\int_0^{t_\mathrm{f}}\rmd s \dot \l(\l - x)\right\rangle\nonumber\\
    =& \kappa\int_0^{t_\mathrm{f}}\rmd s (\dot \l_0+\sigma \dot \l_1)(\l_0+\sigma\l_1 - r_0-\sigma r_1)\nonumber\\
    =& \underbrace{\kappa\int_0^{t_\mathrm{f}}\rmd s \dot \l_0(\l_0- r_0)}_{\equiv W_0} \nonumber\\&+ \sigma\underbrace{\kappa\int_0^{t_\mathrm{f}}\rmd s\left[\dot \l_1(\l_0-r_0) + \dot \l_0(\l_1-r_1)\right]}_{\equiv W_1}\,.
\end{align}
We know from Eq.~\eqref{ConservativeWork} that the 0th order term can be rewritten as
\begin{align}
    W_0 = \frac{\kappa}{2}[\l_0(s)-r_0(s) ] + V(r_0(s))\Big|_{s=0}^{s=t_\mathrm{f}} + \int_0^{t_\mathrm{f}}
    \rmd s\,
    \dot r_0^2\,.\label{appZeroNoiseWork}
\end{align}

Performing a partial integration step, we can express $W_1$ as
\begin{align}
    W_1 =& \kappa\left[ \l_1(\l_0-r_0) +  \l_0(\l_1-r_1)\right]\Big|_{s=0}^{s=t_\mathrm{f}}\nonumber\\
    &-\kappa\int_0^{t_\mathrm{f}}\rmd s\left[ \l_1(\dot\l_0-\dot r_0) + \l_0(\dot \l_1-\dot r_1)\right]\nonumber\\
    =& \kappa\left[ \l_1\left(\l_0-r_0\right) -  \l_0r_1\right]\Big|_{s=0}^{s=t_\mathrm{f}}+\kappa\int_0^{t_\mathrm{f}}\rmd s\left[ \l_1\dot r_0 + \l_0\dot r_1\right]\,.
\end{align}
Next, we substitute $\l_0$ and $\l_1$. To this end, we rearrange~\eqref{AppNoiseExr0} and~\eqref{AppNoiseExr1}, which yields
\begin{align}
    \l_0 &= \frac{1}{\kappa}\left[\dot r_0 - \Fr\right] + r_0\,,\\
    \l_1 &= \frac{1}{\kappa}\left[\dot r_1 - r_1\pr\Fr\right] + r_1\,,
\end{align}
which we then insert into $W_1$ to obtain
    \begin{align}
    W_1 =& \kappa\left[ \l_1\left(\l_0-r_0\right) - \l_0r_1\right]\Big|_{s=0}^{s=t_\mathrm{f}}\nonumber\\
    &+\kappa\int_0^{t_\mathrm{f}}\rmd s\left\{\frac{1}{\kappa}\left[\dot r_1 - r_1\pr\Fr\right] + r_1\right\}\dot r_0 \nonumber\\
    &+ \kappa\int_0^{t_\mathrm{f}}\rmd s\left\{\frac{1}{\kappa}\left[\dot r_0 - \Fr\right] + r_0\right\}\dot r_1\nonumber\\
    =& \kappa\left[ \l_1\left(\l_0-r_0\right) -  \l_0r_1+r_0r_1\right]\Big|_{s=0}^{s=t_\mathrm{f}}\nonumber\\
    &+\int_0^{t_\mathrm{f}}\rmd s\left[\dot r_1 - r_1\pr\Fr\right] \dot r_0 \nonumber\\
    &+ \int_0^{t_\mathrm{f}}\rmd s\left[\dot r_0 - \Fr\right] \dot r_1\,.
\end{align}
Using that
\begin{align}
    \frac{\rmd }{\rmd s}[r_1\Fr] = \dot r_1\Fr + r_1\pr\Fr\dot r_0
\end{align}
we finally find
\begin{align}
    W_1 =& \Big[\kappa 
    \left(\l_0- r_0\right)\left(\l_1- r_1\right) 
    -r_1\Fr\Big]\Big|_{s=0}^{s=t_\mathrm{f}}\nonumber\\&+\int_0^{t_\mathrm{f}}\rmd s2\dot r_1  \dot r_0\,.\label{appsigmaNoiseWork}
\end{align}

The effective Lagrangians entering the work
\begin{align}
    W = \text{boundary term} + \int_0^{t_\mathrm{f}}\rmd s \mathcal{L}
\end{align}
can now be read off Eqs.~\eqref{appZeroNoiseWork} and~\eqref{appsigmaNoiseWork}:
\begin{align}
    \mathcal{L} = \mathcal{L}_0 + \sigma \mathcal{L}_1 = \dot r_0^2 + 2\sigma\dot r_0\dot r_1\,.
\end{align}
The corresponding Euler-Lagrange equations are
\begin{align}
    \ddot r_0 &= 0\tag{\text{$r_1$ variation}}\,,\\
    \ddot r_0&=-\sigma\ddot r_1\tag{\text{$r_0$ variation}}\,.
\end{align}
These imply that $r_0(s) = r_0(0) + c_0 s$ and $r_1(s) = r_1(0) + c_1 s$
and the work functional is
\begin{align}
    W(c_0,c_1) =& \Big\{\frac{\kappa}{2}\left[\l_0(s)-r_0(s)\right]^2+V[r_0(s)]\nonumber\\&+ \sigma\kappa\left[\l_0(s)-r_0(s)\right]\left[\l_1(s)-r_1(s)\right]\nonumber\\&-\sigma r_1(s)F[r_0(s)]\Big\}\Big|_0^{t_\mathrm{f}}
    \nonumber\\&+c_0^2t_\mathrm{f} + 2\sigma c_0c_1t_\mathrm{f}\,.
\end{align}
From our assumptions, we know that $c_0$ minimizes $W_0$, i.e., that
\begin{align}
    -\kappa t_\mathrm{f}[\l_0^\mathrm{f} - c_0t_\mathrm{f} - r_0(0)]-t_\mathrm{f} F[r_0(0) + c_0t_\mathrm{f}] + 2t_\mathrm{f} c_0 = 0\,,
\end{align}
and therefore
\begin{align}
    0=\frac{1}{\sigma}\frac{\partial W}{\partial c_1}=&-\kappa t_\mathrm{f}[\l_0^\mathrm{f}- c_0t_\mathrm{f} - r_0(0)]
    \nonumber \\
    & 
     -t_\mathrm{f} F[r_0(0) + c_0t_\mathrm{f}] + 2t_\mathrm{f} c_0
\end{align}
is always satisfied. Additionally, we have that
\begin{align}
    0=\frac{1}{t_\mathrm{f}}\frac{\partial W_1}{\partial c_0}=& -\kappa [\l_1^\mathrm{f}-c_1t_\mathrm{f} - r_1(0)]+2c_1\nonumber\\&-[c_1 t_\mathrm{f} + r_1(0)]\pr F[r_0(0) + c_0t_\mathrm{f}]\,,
\end{align}  
which we can rearrange to get
\begin{align}
    \left\{\kappa t_\mathrm{f} + 2 -t_\mathrm{f}\pr F[r_0(t_\mathrm{f})]\right\}c_1=&
    r_1(0)\pr F[r_0(t_\mathrm{f})]
    \nonumber\\&+ 
    \kappa [\l_1^\mathrm{f} - r_1(0)]
    \label{appC1lownoise}\,.
\end{align}
Recall that the boundary conditions for $\l$ are $\l(0) = \l^0$ and $\l(t_\mathrm{f})= \l^\mathrm{f}$. If $\l_0$ is to minimize $W_0$, then $\l_0^\mathrm{f}=\l^\mathrm{f}$ and $\l_0^0=\l^0$. Moreover, if $r(0)$ is the initial condition for the particle, then $r_0(0)=r(0)$. As a consequence, $\l_1^\mathrm{f}=0=\l_i^0$ and $r_1(0)=0$. Therefore, using Eq.~\eqref{appC1lownoise}, it follows that $c_1=0$ for all potentials $V(x)$. 

\subsection{Expansion of Work to Second Order}
Next, we go one order further to the second order in the work expansion. Note that equality now is to be understood as ``equal up to $\mathcal{O}(\sigma^3)$''. We proceed in a similar way as the previous section and recognize that
\begin{align}
    W = W_0 + \sigma W_1 + \sigma^2W_2\,,
\end{align}
with $W_0$ and $W_1$ being given by Eqs.~\eqref{appZeroNoiseWork} and~\eqref{appsigmaNoiseWork}, respectively, while
\begin{align}
    W_2 = \kappa\int_0^{t_\mathrm{f}}\rmd s\left\{\dot\l_0(\l_2-r_2)+\dot\l_1(\l_1-r_1)+\dot\l_2(\l_0-r_0)\right\}\,.
\end{align}
Integrating by parts and identifying the boundary contributions yields
\begin{align}
    W_2 =& \kappa \left[\l_0(\l_2-r_2)+\l_1(\l_1-r_1)+\l_2(\l_0-r_0)\right]\Big|_0^{t_\mathrm{f}}\nonumber\\
    &-\kappa\int_0^{t_\mathrm{f}}\rmd s\left[\l_0\left(\dot\l_2- \dot r_2\right)+\l_1\left(\dot\l_1-\dot r_1\right)+\l_2\left(\dot\l_0-\dot r_0\right)\right]\nonumber\\
    =& \kappa \left[\l_0\left(\l_2-r_2\right)+\l_1\left(\frac{1}{2}\l_1-r_1\right)-\l_2r_0\right]\Big|_0^{t_\mathrm{f}}\nonumber\\
    &+\underbrace{\kappa\int_0^{t_\mathrm{f}}\rmd s\left(\l_0\dot r_2+\l_1\dot r_1+\l_2\dot r_0\right)}_{\equiv I}\,.
\end{align}
We now rewrite $I$ by rearranging the average equation of motion for $\lambda_i$ and inserting these into $I$
\begin{align}
I =&\kappa\int_0^{t_\mathrm{f}}\rmd s\Big(\left\{\frac{1}{\kappa}[\dot r_0 -\Fr] +r_0\right\}\dot r_2\nonumber\\&+\left\{\frac{1}{\kappa}\left[\dot r_1 -r_1\pr\Fr\right] +r_1\right\}\dot r_1\nonumber\\&+\left\{\frac{1}{\kappa}\left[\dot r_2 -r_2\pr\Fr -\frac{\langle x_1^2\rangle}{2}\prt\Fr\right] +r_2\right\}\dot r_0\Big)\nonumber\\
=& \left[\kappa r_0r_2+\frac{\kappa}{2}r_1^2-r_2\Fr\right]\Big|_0^{t_\mathrm{f}} + \int_0^{t_\mathrm{f}}\rmd s \Big\{2\dot r_0\dot r_2 \nonumber\\&+ \dot r_1^2 - r_1\dot r_1\pr\Fr - \frac{\langle x_1^2\rangle}{2}\prt\Fr\dot r_0\Big\}\,.
\end{align}  
To simplify this equation, we use $\mathrm{var}(x_1)\equiv \langle x_1^2\rangle - \langle x_1\rangle^2=\langle x_1^2\rangle - r_1^2$, so that, $\frac{\rmd }{\rmd t}\langle x_1^2\rangle = 2\dot r_1 r_1 + \frac{\rmd}{\rmd t}\mathrm{var}(x_1)$. Then, we can rewrite:
\begin{widetext}
\begin{align}
    r_1\dot r_1\pr\Fr + \frac{\langle x_1^2\rangle}{2}\prt\Fr\dot r_0 =& \Bigg[\underbrace{r_1\dot r_1 + \frac{1}{2}\frac{\rmd}{\rmd t}\mathrm{var}(x_1)}_{=\frac{\rmd}{\rmd t}\langle x_1^2\rangle/2}- \frac{1}{2}\frac{\rmd}{\rmd t}\mathrm{var}(x_1)\Bigg]\pr\Fr  +\frac{\langle x_1^2\rangle}{2}\underbrace{\prt\Fr\dot r_0}_{=\frac{\rmd}{\rmd t}\pr\Fr}\nonumber\\
    =&\frac{\rmd }{\rmd t}\left[\frac{\langle x_1^2\rangle}{2}\pr\Fr\right]-\frac{1}{2}\left[\frac{\rmd}{\rmd t}\mathrm{var}(x_1)\right]\pr\Fr\,.
\end{align}
\end{widetext}
Hence, the integral contribution $I$ simplifies to
\begin{align}
    I =& \int_0^{t_\mathrm{f}}\rmd s \left[2\dot r_0\dot r_2 + \dot r_1^2 +\frac{1}{2}\pr\Fr\frac{\rmd}{\rmd t}\mathrm{var}(x_1)\right] \nonumber\\&+ \text{boundary terms}\,.
\end{align}
Hence, the overall Lagrangian is
\begin{align}
    \mathcal{L} =&~ \mathcal{L}_0+\sigma\mathcal{L}_1+\sigma^2\mathcal{L}_2\nonumber\\
    =&~\dot r_0^2 + 2\sigma \dot r_1\dot r_0 + 2\sigma^2\dot r_0\dot r_2 + \sigma^2\dot r_1^2 \nonumber\\&+ \frac{\sigma^2}{2}\left[\frac{\rmd}{\rmd t}\mathrm{var}(x_1)\right]\pr\Fr\label{LagrangeSigma2}\,.
\end{align}
From Eq.~\eqref{x1Expansion}, the solution for $x_1$ can be written as
    \begin{align}
        x_1(t) =&~ x_1(0)\Gamma_-(t) + \kappa \Gamma_-(t)\int_0^{t}\rmd s\lambda(s)\Gamma_+(s)
        \nonumber
        \\
        &+\Gamma_-(t)\int_{s=0}^{s=t}\Gamma_+(s)\rmd W_s\,,
    \end{align}
with
\begin{align}
    \Gamma_\pm(t) = \mathrm{e}^{\pm\int_0^t[\kappa-\pr F(r_0(s))]\rmd s}\,.
\end{align}
Therefore,
\begin{align}
    \mathrm{var}(x_1)(t) &= \Gamma_-(t)^2\int_0^t\Gamma_+(s)^2\rmd s\nonumber\\
    &=\int_0^t \mathrm{e}^{-2\int_z^t[\kappa-\pr F(r_0(s))]\rmd s}\rmd z\,,
\end{align}
and thus
\begin{align}
    \frac{\rmd}{\rmd t}\mathrm{var}(x_1)(t) =& 2\underbrace{\mathrm{e}^{-\int_t^t[\kappa-\pr F(r_0(s))]\rmd s}}_{=1} \nonumber\\&-2[\kappa-\pr F(r_0(t))]\mathrm{var}(x_1)(t)\,.
\end{align}
That is, the only dependence on $r_0$ in the Lagrangian is through the last term in Eq.~\eqref{LagrangeSigma2}, which is a functional of $\pr\Fr$. This term will for non-harmonic potentials lead to nonlinear and nonlocal Euler-Lagrange equations, giving rise to possibly nonlinear  corrections of the colloid trajectory that strongly depend on the specific form of the interaction forces and external fields.

\subsection{Consistency Check: Harmonic Interactions}
To check the consistency of our expansion, now consider the case where all potentials in the system are purely harmonic, i.e., of the form $V(x) = ax^2+bx+c$. Clearly, the potential force then is $F(x) = -2ax - b$ and the only dependence on $r_0$ entering through $\partial_{r_0} F(r_0) = -2a=\mathrm{const.}$ is constant in $r_0$. Hence, the only contribution from a variation in $r_0$ of the work comes from the temporal derivative $\dot r_0$, since variations of $\pr F(r_0)$ yield terms proportional to $\prt \Fr$. The Euler Lagrange equations therefore read
\begin{align}
    2\ddot r_0 + 2\sigma\ddot r_1 + 2\sigma^2\ddot r_2 &= 0\,,\\
    2\sigma\ddot r_0 + 2\sigma^2\ddot r_1&=0\,,\\
    2\sigma^2\ddot r_0&=0\,,
\end{align}
which yield linear solution at every order considered. The exact slope of these linear solution are then determined by the boundary condition of the trap and initial condition for the colloid. However, the initial and final conditions are fixed and satisfied by the $\sigma^0$ solution, as it needs to be consistent with the $\sigma\to0$ limit. Hence, the optimal protocol to second order in $\sigma$ is indistinguishable from the noise free solution. In fact, it is easy to show that this generalizes to higher orders of $\sigma$ as well, since the Lagrangian picks up terms of functionals of $\pr\Fr$ or derivatives thereof. In other words, for any noise amplitude the optimal protocol will be indistinguishable from the noiseless case. We would like to emphasize that this is consistent with the discussion in Ref.~\cite{optimal2007schmiedl}, in which the authors state that the solution for displacing the trap with an overdamped colloid is valid for any noise strength. Our result is slightly more general, as we include not only interactions of the colloid with the trap, but also with an external potential $V(x)$.

\section{Optimal Control of two Particles Coupled by a Spring}\label{AppendixExampleSpring}

Here, we present the detailed calculations for the optimal control protocols for two harmonically coupled overdamped particles with interaction potential $V_\mathrm{int}(\fx_1,\fx_2) = \frac{\Omega}{2}\left[\norm{\fx_1-\fx_2}-l\right]^2$ with rest length $l\geq0$ and coupling strength $\Omega$. We will proceed in three steps of increasing complexity: (i) motion along the coupling force between the particles, (ii) motion of anti-parallel trap displacement with vanishing rest length $l=0$, and (iii) anti-parallel displacement with finite rest length $l>0$. The equations of motion for the particles are given by the overdamped Langevin equation,
\begin{align}
    \f \gamma \rmd{\fx}_{1,s} =& - \kappa[\fx_{1,s}-\fl_1(s)]\rmd s \nonumber\\&- \Omega\frac{\norm{\fx_{1,s}-\fx_{2,s}} - l}{\norm{\fx_{1,s}-\fx_{2,s}}}[\fx_{1,s}-\fx_{2,s}]\rmd s \nonumber\\&+ \sqrt{2k_\mathrm{B}T\f\gamma}\,\rmd W_{1,s},\label{AppLangevin1OD}\\
    \f \gamma \rmd{\fx}_{2,s} =& - \kappa[\fx_{2,s}-\fl_2(s)]\rmd s \nonumber\\&+ \Omega\frac{\norm{\fx_{1,s}-\fx_{2,s}} - l}{\norm{\fx_{1,s}-\fx_{2,s}}}[\fx_{1,s}-\fx_{2,s}]\rmd s \nonumber\\&+ \sqrt{2k_\mathrm{B}T\f\gamma}\,\rmd W_{2,s},\label{AppLangevin2OD}
\end{align}
where 
$\kappa$ is the strength of the optical trap, $V_{\mathrm{t},i}=\frac{\kappa}{2}\norm{\fx_i-\fl_i}^2$, and $\f \gamma = \gamma \mathbf{I}$ is the friction matrix. The force field acting on the colloids are conservative and may be written as the corresponding gradients of the total potential energy $V=V_\mathrm{t,1} + V_\mathrm{t,2} + V_\mathrm{int}$. 
For finite rest length $l>0$, the Langevin Eqs.~(\ref{AppLangevin1OD}, \ref{AppLangevin2OD}) are generally nonlinear, so that the noise average cannot easily be performed straightforwardly. In the following, we therefore focus on analytically tractable special cases: (i) situations in which linearity is restored---either for finite $l>0$ when the displacement occurs only along the force axis, or for zero rest length $l=0$ for arbitrary configuration---and (ii) the athermal cases ($T=0$) with arbitrary configuration and $l\geq 0$.

\subsection{Motion Along Coupling Force Axis}
In this section, we consider the case where the initial and final trap positions are on the axis spanned by the initial harmonic force vector between the two colloids, i.e., $\fr_2(0)-\fr_1(0)$ is parallel to $\fl_1(0)-\fl_1(t_\mathrm{f})= \fl_2(t_\mathrm{f})-\fl_2(0)$. This leaves two options, moving the traps either towards or away from each other. Due to the spatial symmetry of such motion, we expect no average motion perpendicular to the force axis (any noise component perpendicular to the force axis will, with equal probability, have an equally strong opposite noise component). Hence, we can use an effectively one-dimensional model. Thus, we drop the bold font, as positions and velocities are scalar quantities along the force axis, 
and the Euclidean norm simplifies $\norm{x_1-x_2} = |x_1-x_2|$. Without loss of generality, we 
chose the axis to be such that $r_2(0) = -r_1(0)>0$ and $\lambda_2(0) = -\lambda_1(0)\equiv\lambda^0>0$. The problem we pose has the same total displacement of the traps, hence $\lambda_2(t)=-\lambda_1(t)\equiv\lambda^\mathrm{f}\geq 0$. As a consequence, we have that $\norm{R_1(s)-R_2(s)} = R_2(s) - R_1(s)$ for all $s$ making the system linear. The average equation of motion read
\begin{align}
    \gamma \dot{r}_1(s) &= -\kappa[r_1(s)-\lambda_1(s)] - \Omega[r_1(s)-r_2(s) + l]\,,\\
    \gamma \dot{r}_2(s) &= -\kappa[r_2(s)-\lambda_2(s)] - \Omega[r_2(s)-r_1(s) - l]\,.
\end{align}
We can further simplify the problem by changing to relative coordinates $r(s) \equiv r_2(s) - r_1(s)$ and center of mass coordinates $z(s)\equiv r_2(s) + r_1(s)$, in which the equations of motion read
\begin{align}
    \gamma \dot{r}(s) &= -\kappa[r(s) - \lambda_r(s)] - 2\Omega[r(s) - l]\label{1dSpringRel},\\
    \gamma\dot{z}(s) &= -\kappa[z(s) - \lambda_z(s)]\label{1dSpringCent},
\end{align}
where $\lambda_r(s) = \lambda_2(s) - \lambda_1(s)$ and $\lambda_z(s) = \lambda_2(s) + \lambda_1(s)$. Using that $2[(r_1 - \lambda_1)^2 +(r_2-\lambda_2)^2] = (r-\lambda_r)^2 + (z-\lambda_z)^2$, we get
\begin{align}
    2W[\lambda_r, \lambda_z] =& \frac{\kappa}{2}\left[[\lambda_r(s) - r(s)]^2 + [\lambda_z(s) - z(s)]^2\right]_0^{t_\mathrm{f}}\nonumber\\
    &+\Omega[r(s) - L]^2\Big|_0^{t_\mathrm{f}}+\int_0^{t_\mathrm{f}}\rmd s\left[\dot{r}(s)^2 + \dot{z}(s)^2\right],
\label{1DSpringWorkFunctional}
\end{align}
by rearranging and inserting Eqs.~\eqref{1dSpringRel} and~\eqref{1dSpringCent} for $\lambda_r$ and $\lambda_z$, respectively. The Euler-Lagrange equations are therefore $\ddot{r}(s) = 0 = \ddot{z}(s)$, with solution $r(s) = r(0) + c_r s$ and $z(s) = z(0) + c_z s$, with unknown parameters $c_r$ and $c_z$. The latter can be determined by plugging in the optimal solutions into the work functional~\eqref{1DSpringWorkFunctional}, and minimizing with respect to $c_r$ and $c_z$. Since the $z$ and $r$ coordinate decouple, we may treat these minimizations separately. Optimizing $W$ w.r.t. $c_z$, we find that
\begin{align}
    \!\!\!W(c_z) = \mathrm{const.} + \frac{\kappa}{2}\left[\lambda_z({t_\mathrm{f}}) - z(0) - c_z {t_\mathrm{f}}\right]^2 + \gamma c_z^2 {t_\mathrm{f}}
\end{align}
is minimized by
\begin{align}
    c_z = \frac{\kappa}{\kappa {t_\mathrm{f}}+2\gamma}\left[\lambda_z({t_\mathrm{f}}) - z(0)\right]\,.
\end{align}
Since, $\lambda_z({t_\mathrm{f}}) = z(0)$, we thus find $c_z=0$. Similarly, optimizing $W$ w.r.t. $c_r$, we find that
\begin{align}
    W(c_r) =& \mathrm{const.} + \frac{\kappa}{2}\left[\lambda_r({t_\mathrm{f}}) - r(0) - c_r {t_\mathrm{f}}\right]^2 \nonumber\\&+ \gamma c_r^2 {t_\mathrm{f}} + \Omega\left[r(0) + c_r {t_\mathrm{f}} - l\right]^2
\end{align}
is minimized by
\begin{align}
    c_r = \frac{\kappa[\lambda_r({t_\mathrm{f}}) - r(0)] - 2\Omega \left[r(0)-l\right]}{{t_\mathrm{f}}(\kappa+2\Omega) + 2\gamma}\,.\label{1dSpringRelSlope}
\end{align}
From Eq.~\eqref{1dSpringRel}, we then get
\begin{align}
    r(0) = \frac{2\Omega l + \kappa\lambda_r(0)}{2\Omega + \kappa} = \frac{2\Omega l + 2\kappa\lambda^0}{2\Omega +\kappa},
\end{align}
which we can insert into Eq.~\eqref{1dSpringRelSlope} to get
\begin{align}
    c_r &= \frac{2\kappa\lambda^\mathrm{f} +2\Omega l - r(0)\left(\kappa + 2\Omega\right)}{{t_\mathrm{f}}(\kappa+2\Omega) + 2\gamma}\nonumber\\
    &=\frac{2\kappa}{{t_\mathrm{f}}(\kappa+2\Omega) + 2\gamma}(\lambda^\mathrm{f} - \lambda^0)\,.
\end{align}
Thus, the optimal relative motion of colloid and trap are
\begin{align}
    r(s)=&\frac{2\Omega l+2\kappa\lambda^0}{\kappa+2\Omega} + \frac{2\kappa}{{t_\mathrm{f}}(\kappa+2\Omega) + 2\gamma}(\lambda^\mathrm{f} - \lambda^0)s\\
    \lambda_r(s) =& 2\frac{(\kappa+2\Omega)s+\gamma}{(\kappa+2\Omega){t_\mathrm{f}}+2\gamma}\lambda^\mathrm{f} + 2\frac{(\kappa+2\Omega)({t_\mathrm{f}}-s)+\gamma}{(\kappa+2\Omega){t_\mathrm{f}} + 2\gamma}\lambda^0\,,
\end{align}
which corresponds to the well known one-dimensional single particle solution from Ref.~\cite{optimal2007schmiedl} with effective $k_\mathrm{eff} = \kappa+2\Omega$ and $\lambda^0\neq 0$. The individual trap protocols can be obtained from this solution using $\lambda_2(s) = [\lambda_r(s) - \lambda_z(s)]/2 = \lambda_r(s)/2$ and $\lambda_1(s) = -\lambda_r(s)/2$.

\subsection{2D Configuration with Vanishing Rest Length}

Next, we turn to the case with vanishing rest length, $l=0$, where the traps displacement is not aligned with the spring force axis. As a specific example, we consider the counter propagating case with $\fl_1(0) = (0,0)^T$ to $\fl_1({t_\mathrm{f}})=(0,\He)^T$ and $\fl_2(0)=(\Dx, \He)^T$ to $\fl_2({t_\mathrm{f}})=(\Dx, 0)^T$, where $\Dx,\He>0$ are horizontal and vertical displacements, sketched in Fig.~\ref{fig:SpringLengthComparison}(a). The choice $l=0$ renders Eqs.~\eqref{AppLangevin1OD} and~\eqref{AppLangevin2OD} linear, so that we can perform the noise average directly (for any $T$). In other words, we know that Eq.~\eqref{ConservativeWork} is valid and thus the optimal colloid trajectories are $\fr_i(s)=\fr_i(0)+\f c_i s$. Since $l=0$, the horizontal and vertical components of the slope vectors $\f c_i$ decouple in the work
\begin{align}
    W =&\sum_{i=1}^2\frac{\kappa}{2}\norm{\fl_i(s) - \fr_i(0)-\f c_i s}^2\Big|_0^{t_\mathrm{f}}+{t_\mathrm{f}}\sum_{i=1}^2\f c_i\cdot\f\gamma\f c_i \nonumber
    \\&
    + \frac{\Omega}{2}\norm{\fr_1(0)+\f c_1 s-\fr_2(0) - \f c_2 s}^2\Big|_0^{t_\mathrm{f}}\,,
\end{align}
which leads to the minimization conditions
\begin{align}
    \f c_1 &= \frac{\kappa\fl_1({t_\mathrm{f}}) - (\kappa+\Omega)\fr_1(0) +\Omega\fr_2(0) + \Omega {t_\mathrm{f}}\f c_2}{{t_\mathrm{f}}(\kappa+\Omega) + 2\gamma}\,, \\
    \f c_2 &= \frac{\kappa\fl_2({t_\mathrm{f}}) - (\kappa+\Omega)\fr_2(0) +\Omega\fr_1(0) + \Omega {t_\mathrm{f}}\f c_1}{{t_\mathrm{f}}(\kappa+\Omega) + 2\gamma}\,,
\end{align}
where we use that $\f \gamma = \gamma \mathbf{I}$. Using that the initial system is in equilibrium and hence $\dot \fr_i(0)=0$, we can solve the average Langevin equations, obtaining
\begin{align}
    \fr_1(0) &= \frac{\kappa\fl_1(0) + \Omega\fr_2(0)}{\kappa+\Omega}\,, \\
    \fr_2(0) &= \frac{\kappa\fl_2(0) + \Omega\fr_1(0)}{\kappa+\Omega}\,,
\end{align}
so that
\begin{align}
    \fr_1(0) &= \frac{(\kappa+\Omega)\fl_1(0) + \Omega\fl_2(0)}{\kappa+2\Omega}\,, \\
    \fr_2(0) &= \frac{(\kappa+\Omega)\fl_2(0) + \Omega\fl_1(0)}{\kappa+2\Omega}\,.
\end{align}
Using that $\fl_1({t_\mathrm{f}})-\fl_1(0) = -[\fl_2({t_\mathrm{f}}) - \fl_2(0)] = (0, \He)^T$, the slopes become
\begin{align}
    \f c_1 &= \frac{\kappa}{{t_\mathrm{f}}(\kappa+2\Omega) + 2\gamma}\begin{pmatrix}
        0\\ \He
    \end{pmatrix}\,, \\
    \f c_2 &= -\frac{\kappa}{{t_\mathrm{f}}(\kappa+2\Omega) + 2\gamma}\begin{pmatrix}
        0\\ \He
    \end{pmatrix}\,,
\end{align}
which has the same characteristics as the 1D solution. Notably, {since the slope of $\fr_i$ is $\f c_i$}, {the slope of }the optimal protocols {are also} proportional to $\f c_1$, so one never benefits from horizontal displacement.

\subsection{2D Configuration with Finite Rest Length}
Lastly, we turn to the counterpropagating case with finite rest length $l>0$. 
The mismatch in force axis and trap displacement makes the problem nonlinear, hence we can only make statements in the athermal limit.  
From our general considerations, we immediately know that the optimal particle trajectories again follow steady straight-line motions. 

To obtain the speeds and slopes, we would need to perform a secondary optimization of the total work functional. The term $\Delta V$, renders this a nontrivial nonlinear problem, which can be solved using numerical tools.

However, in addition to the geometric and temporal linearity of the optimal paths, further symmetry considerations allow us to understand the general structure of the solutions without fully solving it.
Rotating the system by $\pi$ and relabeling $1\leftrightarrow2$ recovers the original system 
As a consequence, the center of mass of the colloids remains at $\fr_\mathrm{cms}=(\Dx/2, \He/2)$. The part of 2D space enclosed by a circle of radius $l/2$ around $\fr_\mathrm{cms}$ we denote by $S_{l/2}=\{(x,y)^T\in\mathds{R}^2|\frac{\Dx-l}{2}\leq x\leq \frac{\Dx+l}{2}, \frac{\He-l}{2}\leq y\leq \frac{\He+l}{2}, x^2+y^2\leq l/2 \}$. Then, by symmetry, if $\fr_1(s)\in S_{l/2}$ for any $s$, then also $\fr_2(s)\in S_{l/2}$ for the same $s$ and therefore $\norm{\fr_1(s)-\fr_2(s)}\leq l$. Hence, we can again use the single colloid picture to effectively describe the dynamics of the system. Clearly, $\fr_1\in S_{l/2}$ means that the colloid is \emph{repulsed} away from $\fr_\mathrm{rms}$, while $\fr_1\notin S_{l/2}$ means that the colloid is \emph{attracted} towards $\fr_\mathrm{cms}$. One can therefore immediately see that nontrivial behaviour arises in the optimal control from a force-balance perspective. If $l$ is sufficiently large, the colloids will enter the ``repulsion disc'' $S_{l/2}$, where the spring force becomes repulsive. As a consequence, the optical trap will have to move to compensate this switch in force direction, as we know that the colloids move in a straight line. Due to the circular geometry of the spring force, there will be a maximal repulsion, followed by a decrease afterwards; hence the optical traps will have to adjust, leading to nonlinear (V-shaped) behaviour.



\nolinenumbers
\newpage

\nolinenumbers
\bibliography{bib}

\end{document}